\documentclass[10pt,
               aps,
               prl,
               twocolumn,
               showpacs,
               groupedaddress]{revtex4-2}
\usepackage[utf8]{inputenc}
\usepackage{graphicx}
\graphicspath{{images/}}
\usepackage[caption=false]{subfig}
\usepackage{amsmath, amsfonts, amssymb}
\usepackage[linktocpage=true,
  colorlinks=true, 
  pdfborder={0 0 0},
  linkcolor=blue,
  citecolor=blue,
  filecolor=yellow,
  urlcolor=blue,
  bookmarks,
  pdfauthor={},
]{hyperref}
\usepackage{tikz}
\makeatletter
\newcommand*{\rom}[1]{\expandafter\romannumeral #1}
\makeatother
\begin{document}

\title{Active magneto gyrator as a tunable heat engine or heat pump}

\author{F. Adersh}


\author{M. Muhsin}

\affiliation{Department of Physics, University of Kerala, Kariavattom, Thiruvananthapuram-$695581$, India}

\author{M. Sahoo}
\email{jolly.iopb@gmail.com}
\affiliation{Department of Physics, University of Kerala, Kariavattom, Thiruvananthapuram-$695581$, India}

\date{\today}

\begin{abstract}
 We theoretically investigate the thermodynamic performance characteristics of an active magneto-gyrator taking into account the two-dimensional motion of an inertial charged active particle confined in an asymmetric parabolic potential and in contact with two heat baths kept at two different temperatures. 
 A magnetic field of constant magnitude is applied in a direction perpendicular to the plane of motion.
 In such a system, the particle exhibits a gyrating motion across the potential minimum and exerts a torque on the confining potential as long as there is a potential asymmetry and temperature gradient. Hence, this system can operate as an active magneto-heat engine or pump in the presence of a load force. Interestingly, we observe that the activity or self-propulsion impacts the thermodynamic performance characteristics of the gyrator only in the presence of the magnetic field. We examine two scenarios: first, by applying a load in a direction opposing the torque and second, by applying a load in the same direction as that of torque. 
 In the first case, for a fixed parameter regime, the gyrator is found to act as a heat engine or a heat pump depending on the strength of the applied load, whereas in the latter case, it can only operate as a heat pump.  
Moreover, unlike the Brownian gyrator or Brownian magneto gyrator, captivatingly, the efficiency is found to have no universal upper bound and can be made $100\%$ by tuning the system parameters. Additionally, when the system is suspended in a viscoelastic medium characterized by the presence of a finite memory, for a short persistence of memory and a fixed duration of activity, the efficiency can be $100\%$ even for more than one value of viscoelastic memory timescale.
In the first case, the duration of the activity has a non-monotonic impact on the system performing either as an engine or a pump, whereas the magnetic field has a similar impact on the performance of an engine, but it degrades the performance of the system as a heat pump. In the latter case, the coefficient of performance has a non-monotonic dependence on the duration of the activity, whereas the magnetic field favors the system performing as a heat pump.

\end{abstract}

\maketitle
\section{INTRODUCTION}\label{sec:intro}
A Brownian gyrator is a fundamental stochastic model consisting of a Brownian particle in an asymmetric harmonic potential, undergoing a gyrating motion across the potential minimum while being coupled to two thermal baths at different temperatures~\cite{filliger2007brownian, dotsenko2013two, mancois2018two, Chiang2017electrical, bae2021inertial, Argun2017Experimental, Chang2021autonomous, baldassarri2020engineered, cerasoli2021time, fogedby2017minimal, cerasoli2022spectral}.
The temperature gradient drives heat flow from the hot bath to the cold bath, thereby pushing the system out of equilibrium. Such systems play a crucial role in non-equilibrium statistical mechanics and stochastic energetics~\cite{murashita2016overdamped, nascimento2020memory, dotsenko2013two, broeck2004micro, berut2016role, crisanti2012nonequilibrium, park2016efficiency}. 
A Brownian gyrator serves as a minimal model to generate a systematic torque, which can be exploited to build a microscopic-scale heat engine~\cite{park2016efficiency, abdoli2022tunable, Chiang2017electrical, Chang2021autonomous}.
Various theoretical models exist for the realization of heat engines from Brownian particles~\cite{broeck2004micro, filliger2007brownian, fogedby2017minimal}, including cyclic heat engines such as Brownian Carnot engines, where Brownian particles operate in time-dependent potentials~\cite{martinez2016carnot, adam2022browniancarnot}. 
Furthermore, Brownian gyrator-based heat engines have also been experimentally demonstrated in recent studies.~\cite{Argun2017Experimental, Chiang2017electrical}.

A constant magnetic field can induce Lorentz force-driven spatial correlations in two-temperature systems~\cite{abdoli2020correlations}. In addition, it can adjust the performance characteristics while acting as a heat engine or heat pump~\cite{abdoli2022tunable}.
The introduction of activity into such systems offers further tunability, as stochastic heat engines with active Brownian dynamics exhibit advantages over their passive counterparts~\cite{kumari2024Stochastic, saha2019stochastic}.
The stochastic thermodynamics of active systems remains an emerging area of research~\cite{Pietzonka2018Entropy,mandal2017entropy, datta2022second, saha2018stochastic, saha2019stochastic, kumari2020stochastic}.
Interestingly, active heat engines do not have a theoretical upper bound on efficiency, and it can be arbitrarily close to one~\cite{pietzonka2019Autonomous}.
Similarly, active systems also exhibit exotic phenomena, such as motility-induced phase separation (MIPS), which can be used to design an active refrigerator capable of localizing its cooling domain~\cite{HechtActive2022}.  When such a two-temperature system of an inertial charged active particle is subjected to an external magnetic field, the system exhibits interesting phenomena, such as memory-induced trapped diamagnetism~\cite{Muhsin2025Active}. In addition, the effective torque exerted by the particle on the potential is affected by the strength of the magnetic field and hence can influence the performance characteristics of the system. 
Thus, it is also interesting to investigate the thermodynamic performance characteristics of the system by utilizing the effective torque.

In this work, we have investigated the performance characteristics of an active magneto-gyrator composed of an inertial charged active particle moving in an external magnetic field. The system is placed in a two-temperature bath under the influence of an asymmetric parabolic potential. Due to the combined effects of the temperature gradient and potential asymmetry, the system exhibits gyration of the particle across the potential minimum, while the presence of finite activity and a magnetic field induces precession across the field~\cite{Muhsin2025Active}. 
In such a system, the particle exerts a non-zero average torque on the confining potential. The direction of the torque is found to be independent of the direction of the external magnetic field.
With this setup, we discuss two different cases. In the first case, an external load is applied in the opposite direction of the torque to extract energy from the system. In the second case, the load is applied in the same direction as the torque.
For the former case, the system is found to function in different operating modes, such as a heat engine or a heat pump, depending on the direction of extracted power and absorbed heat from the thermal bath. Interestingly, in the heat engine regime, the efficiency of the system can reach the maximum possible value ($100\%$) without any upper bound. This challenging situation cannot be achieved in the absence of either magnetic field or activity. We further analyzed the system by suspending it in a viscoelastic medium characterized by a time-dependent friction kernel that decays exponentially with a memory timescale. In this case, for a short persistence of memory and fixed duration of activity in the medium, the heat engine is observed to achieve $100\%$ efficiency even for more than one value of the memory timescale. Additionally, the activity affects the efficiency of the engine only in the presence of a magnetic field. 
 In the latter case, as extraction of work is not possible, the system in this configuration functions as a heat pump.

We have organized the paper in the following way. In section ~\ref{sec:model}, we introduce the model and the parameters of interest. The results and detailed discussion of all the different cases are presented in section ~\ref{sec:result}, followed by a summary in section ~\ref{sec:summary}.

\section{MODEL}\label{sec:model}
We consider the motion of a charged active Ornstein-Uhlenbeck particle of mass $m$,  confined in a two-dimensional (2d) harmonic potential $U(x,y)=\frac{1}{2}k \left(x^2+y^2\right)+\alpha x y$ with a finite asymmetry. Here $k$ is the harmonic constant and $\alpha$ characterizes the asymmetry of the potential. The parameter $\alpha$ has the same dimensionality as of $k$. A constant magnetic field ${\bf B}=B \hat{k}$ is applied perpendicular to the plane of motion of the particle such that the Lorentz force acts perpendicular to both the velocity of the particle and the magnetic field. The underdamped motion of the particle can be described by using the Langevin equation \cite{Muhsin2025Active, muhsin2022inertial, arsha2021velocity, caprini2021inertial}
\begin{equation}
  m\ddot{{\bf r}}(t) = -\gamma \dot{{\bf r}}(t)+|q|\left({{\bf{\dot{r}}}(t)\times{\bf B}}\right) - \nabla U  +{\bf f}^s(t)+{\bf F}_L+ {{\boldsymbol{\xi}}}(t),
    \label{eq:maineqmotion-vector}
\end{equation}
with ${\bf r}(t)=x(t)\hat{i}+y(t)\hat{j}$, being the position vector of the particle and ${\bf F}_L$ being the load force of the form 
\begin{equation}
    {\bf F}_L = -\epsilon( {\bf r}(t) \times \hat{k}).
    \label{eq:F_L}
\end{equation}
This force ${\bf F}_L$ is perpendicular to the radius vector, with the parameter $\epsilon$ representing the strength of the force. ${\bf F}_L$ is a linear, nonconservative force since $\nabla\times {\bf F}_L\neq0$. It generates a torque in the positive or negative $z$-direction depending on the sign of $\epsilon$. The term $\boldsymbol{\xi}(t)$ is a white noise vector with zero average and correlation $\langle\xi_x(t)\xi_y(t')\rangle=2\delta_{xy} \gamma T_i\delta (t-t')$, where $i=(c,h)$ corresponds to the $x$ and $y$ degrees of freedom in the $xy$-plane and $T_{(c,h)}$ are the cold and hot bath temperatures connected to the $x$ and $y$ coordinates respectively.
The term ${\bf f}^s(t)$ represents the active force that is introduced in the dynamics to model the self-propulsion of the particle, and typically it follows the Ornstein-Uhlenbeck (OU) process

\begin{equation}
t_c\dot {\bf f}^s(t) = -{\bf f}^s(t) + f_0 \sqrt{2 t_c} \boldsymbol{\zeta}(t).
\label{eq:noise}
\end{equation}

The noise $\boldsymbol{\zeta}(t)$ in Eq.~\eqref{eq:noise} is a Gaussian white noise with zero average and unit variance. The active force ${\bf f}^s(t)$ has the properties

\begin{equation}
\langle \text{f}^s_x(t) \rangle =0, \qquad  \langle \text{f}^s_x(t)\text{f}^s_y(t') \rangle =f_0^2 \delta_{xy}e^{-\frac{|t-t'|}{t_c}}.
\label{eq:noise_stat}
\end{equation}
In Eqs.~\eqref{eq:noise} and~\eqref{eq:noise_stat}, the timescale $t_c$ can be identified as the activity timescale. It is the time up to the particle exhibits self-propulsion in the medium. The parameter $f_0$ represents the strength of the active force. The average torque exerted by the particle on the confining potential can be written as
\begin{equation}
    \langle \tau \rangle=-|\langle {\bf r}\times \nabla U \rangle|.
    \label{eq:torque_main}
\end{equation}
One can extract the useful work by applying a constant load force ${\bf F}_L$ in a direction opposite to that of torque on the potential. Hence, the average mechanical power  delivered by the system can be calculated as
\begin{equation}
     P=-\langle {\bf F}_L\cdot {\bf \dot{r}}\rangle.
     \label{eq:power_main}
\end{equation}
From the principles of stochastic thermodynamics, the total heat absorbed by the particle from the hot and cold baths, $d\mathcal{Q}_c$ and $d\mathcal{Q}_h$ respectively, are given by~\cite{sekimoto1997langevin, sekimoto2010stochastic}
\begin{equation}
    d{\mathcal{Q}}_i=\Bigr[-\gamma  {\dot{r}}_i(t)+{\text{f}}_i^s(t)+{ {\xi}_i}(t)\Bigr]\circ d{r}_i,
    \label{eq:dQh}
\end{equation}
with $i = (c,h)$. The symbol `$\circ$' denotes multiplication in the Stratonovich sense.
The steady-state average rate of heat that is dissipated from the hot bath can be calculated as
\begin{equation}
    Q_h=\Bigl \langle \frac{d\mathcal{Q}_h}{dt}\Bigr \rangle.
    \label{eq:Q_h_definition}
\end{equation}

Similarly, the heat dissipated from the cold bath is $d\mathcal{Q}_c$, and the steady-state average rate of heat out of the cold bath is given by
\begin{equation}
    Q_c= \Bigl\langle \frac{d\mathcal{Q}_c}{dt} \Bigr\rangle
    \label{eq:Q_c_definition}
\end{equation}
The efficiency $(\eta)$ of such a system 
can be determined by evaluating how effectively the system converts heat from the hot bath ($Q_h$) into useful work. Therefore, $\eta$ can be calculated by comparing the output work or the extracted power ($P$) with the input heat from the hot bath ($Q_h$) as

\begin{equation}
    \eta=\dfrac{P}{Q_h}.
    \label{eq:Eff_main}
\end{equation}

 When $Q_h<0$, heat is absorbed by the hot bath. In such a situation, the system can be operated as a heat pump, and the performance of the heat pump can be determined by evaluating the coefficient of performance (COP). The COP can be calculated by comparing the heat absorbed by the hot bath with the delivered mechanical power \cite{qi2022thermal, chen2024three} as 
\begin{equation}
     COP=\dfrac{Q_h}{P}.
     \label{eq:COP_main}
\end{equation}

\section{RESULTS AND DISCUSSION}\label{sec:result}

\subsection{Applying load in the opposite direction of torque}\label{sec:A}

When the load ${\bf F}_L$ acts in the opposite direction of the torque, Eq.~\eqref{eq:maineqmotion-vector} can be rewritten as

\begin{equation}
     \ddot{{\bf r}}(t) = -\frac{\gamma}{m} \dot{{\bf r}}(t)-\omega_c {\bf \dot{r}}(t)\times\hat{k}-\dfrac{\nabla U}{m} + \frac{{\bf f}^s(t)}{m} - \frac{{\bf F}_L}{m}+ \frac{{{\boldsymbol{\xi}}}(t)}{m},
     \label{eq:Dynamics_OD}
\end{equation}

where $\omega_c=\dfrac{q|B|}{m}$.  The Eq.~\eqref{eq:Dynamics_OD}, along with Eq.~\eqref{eq:noise}, can be converted into a set of three Markovian equations as follows.

\begin{figure}
\centering
\includegraphics[width=0.8\linewidth]{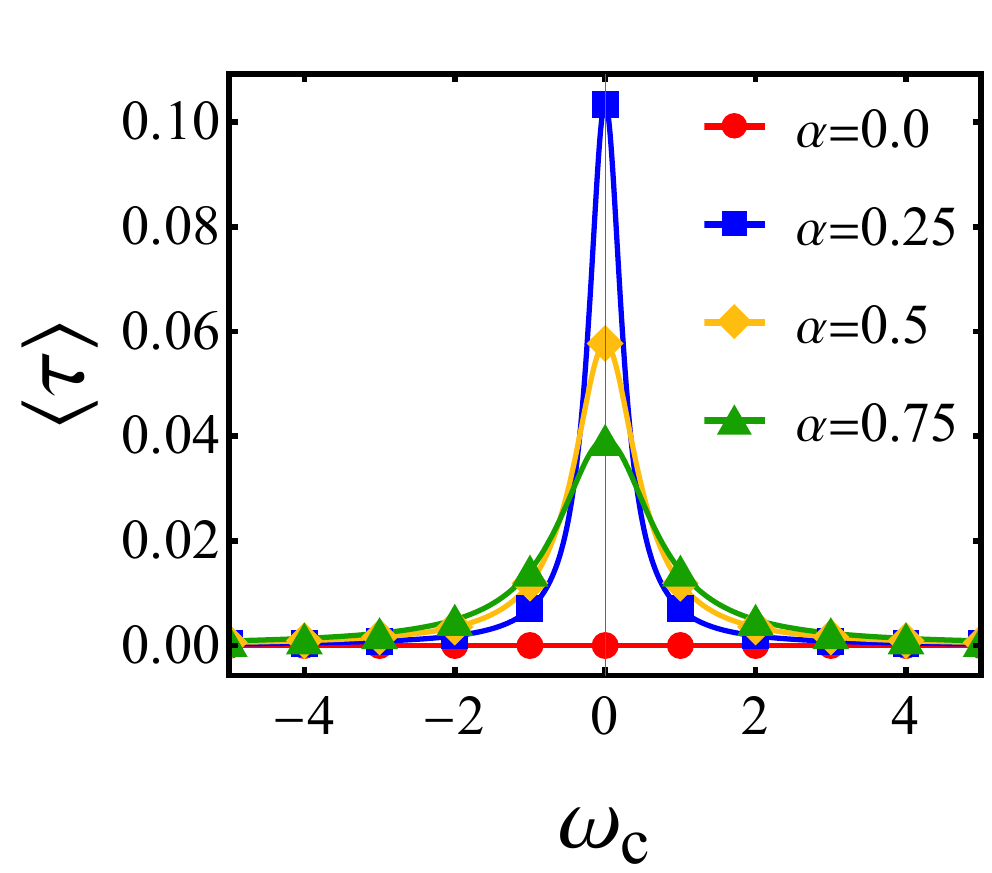}
\caption{ The torque $\langle \tau \rangle$ exerted by the particle on the confining potential $U(x,y)$ as a function of $\omega_c$ for different values of $\alpha$. The other common parameters are $m = f_0 = t_c = \omega_0 =1.0$, $\gamma=0.1$, $T_c=1.0$, and $T_h=4.0$.}
\label{fig:torque_DR}
\end{figure}

\begin{align}
    {\bf \dot{r}} &= {\bf v} \nonumber\\
    {\bf \dot{v}} &= -{\frac{\gamma}{m} \bf v} -\frac{\nabla U}{m} + \left(\omega_c{\bf v} + \frac{\epsilon}{m} {\bf r}\right) \times \hat{k} + \frac{{\bf f}^s}{m} + \frac{{{\boldsymbol{\xi}}}}{m} \\
      {\dot{\bf f}^s} &= -\frac{{\bf f}^s}{t_c} + f_0 \sqrt{\frac{2}{t_c}} \boldsymbol{\zeta}(t)\nonumber.
\end{align}
The above equations can be expressed in matrix form as 
\begin{equation}
    \dot{\boldsymbol{\chi}} = A\boldsymbol{\chi} + B \boldsymbol{\xi'},
    \label{eq:dynamics-matrix}
\end{equation}
where the matrices $\boldsymbol{\chi}$, $A$, $B$ and $\boldsymbol{\xi'}$ are given by
\begin{equation}
    \boldsymbol{\chi} = \begin{pmatrix}
        x & y & v_x & v_y & \xi_x & \xi_y
    \end{pmatrix}^{T},
\end{equation}

\begin{equation}
    A = \begin{pmatrix}
        0 & 0 & 1 & 0 & 0 & 0 \\
        0 & 0 & 0 & 1 & 0 & 0 \\
        -\omega_0^2 & \frac{-\alpha+\epsilon}{m} & -\frac{\gamma}{m} & \omega_c & \frac{1}{m} & 0 \\ 
        \frac{-\alpha-\epsilon}{m} & -\omega_0^2 & -\omega_c & -\frac{\gamma}{m}  & 0 & \frac{1}{m} \\
        0 & 0 & 0 & 0 & \frac{-1}{t_c} & 0 \\
        0 & 0 & 0 & 0 & 0 & \frac{-1}{t_c}
    \end{pmatrix},
\end{equation}

\begin{equation}
    B = \begin{pmatrix}
         0 & 0 & 0 & 0 & 0 & 0 \\
         0 & 0 & 0 & 0 & 0 & 0 \\
         0 & 0 &\frac{\sqrt{2 \gamma T_c}}{m} & 0 & 0 & 0 \\
         0 & 0 & 0 & \frac{\sqrt{2 \gamma T_h}}{m} & 0 & 0 \\
         0 & 0 & 0 & 0 &\sqrt{\frac{2 f_0^2}{t_c}}  & 0 \\
         0 & 0 & 0 & 0 & 0 & \sqrt{\frac{2 f_0^2}{t_c}} \\
    \end{pmatrix},
\end{equation}

and 
\begin{equation}
    \boldsymbol{\xi'} = (0\ 0\ \xi_x\ \xi_y\ \ \zeta_x\ \ \zeta_y)^T.
\end{equation}

To study the steady-state properties of the system, we introduce the correlation matrix $\boldsymbol{\Xi}$ with elements given by
\begin{equation}
    \Xi_{i,j} = \langle \chi_i \chi_j \rangle - \langle \chi_i \rangle \langle \chi_j \rangle
    \label{eq:corr_mat_def}
\end{equation}

As per the correlation matrix formalism \cite{van2007stochastic}, the matrix $\boldsymbol{\Xi}$ can be shown to satisfy the equation
\begin{equation}
    A\cdot \boldsymbol{\Xi} + \boldsymbol{\Xi} \cdot A^T + B\ B^T = 0.
    \label{eq:Xi_relation}
\end{equation}

By solving the Eq.~\eqref{eq:Xi_relation} one can obtain $\boldsymbol{\Xi}$, with which all the physical quantities in Eqs.~\eqref{eq:torque_main}-\eqref{eq:COP_main} can be calculated.
The particle exerts a systematic torque on the confining potential in the presence of a temperature gradient ($T_h \neq T_c$) and a potential asymmetry. Using Eq.~\eqref{eq:torque_main}, this average torque can be calculated as

\begin{equation}
    \langle \tau \rangle = \alpha \left( \Xi_{1,1} - \Xi_{2,2} \right) = \frac{\alpha  \gamma ^2  \left(T_h-T_c\right)}{m\left[\alpha ^2+\omega _0^2 \left(\gamma ^2+m^2 \omega _c^2\right)\right]}.
    \label{eq:torque}
\end{equation}

 In Fig.~\ref{fig:torque_DR}, we have plotted $\langle \tau \rangle$ as a function of $\omega_c$ for different values of $\alpha$ and in the absence of load ($\epsilon = 0$).
 It is observed that $\langle\tau\rangle$ is a decreasing function of $\omega_c$ and the direction of the torque is independent of the applied field. Also, in the $\omega_{c} \to 0$ limit, it varies non-monotonically with $\alpha$, whereas it enhances with $\alpha$ for a finite field strength. It is to be noted that there is no average preferential direction of motion of the particle when there is no asymmetry in potential. Therefore, the average torque is zero when $\alpha=0.0$. Moreover, $\langle \tau \rangle$ is independent of the activity timescale $t_c$, as evident from Eq.~\eqref{eq:torque}.
 
From Fig.~\ref{fig:torque_DR}, it is observed that torque is always positive for positive $\alpha$ values. Therefore, one can extract work from the system by loading the system with a linear force such that it opposes the effective torque. Since the torque is positive, the sign of the applied load $({\bf F}_L)$ should be chosen such that it can exert a torque in the negative $z$-direction. Using Eqs.~\eqref{eq:power_main} and ~\eqref{eq:F_L}, the power generated by the system can be calculated as

\begin{widetext}
    \begin{equation}  
    \begin{split}
P&=\epsilon\langle x\dot{y}-y \dot{x}\rangle\\
&=\Biggr[2 f_0^2 m^2 \epsilon  t_c \omega _c\biggl\{\dfrac{ t_c^2 \left(\alpha ^2 \gamma +\gamma ^3 \omega _0^2-\epsilon \omega _c \left(\gamma ^2+m^2 \omega _0^2\right)+\gamma  m^2 \omega _0^2 \omega _c^2\right)-m^2 \left(\epsilon \omega _c-2 \gamma  m \epsilon  t_c \omega _c\right)}{S}\biggl\}\\
&+\dfrac{2 \gamma ^2 f_0^2 \epsilon ^2 t_c \bigl(m^2 \omega _0^2 t_c^2+\left(\gamma  t_c+m\right)^2 \bigl)}{S}\Biggr]+ \dfrac{\gamma  \epsilon\Delta_0}{m}\left(\dfrac{\epsilon  \left(T_c+T_h\right) \left(\gamma ^2+m^2 \omega _c^2\right)+\alpha  \gamma ^2 \left(T_c-T_h\right)}{S}\right),
\label{eq:Power_FE_OD}
\end{split}
    \end{equation}
where
    \begin{equation}
      S= \Delta _0 \left[\left(\gamma ^2 \omega _0^2+\gamma  \epsilon  \omega _c-\epsilon ^2\right) \left(\gamma ^2+m^2 \omega _c^2\right)-\alpha ^2 \gamma ^2\right],
    \end{equation}

    \begin{equation} 
        \Delta_0=  m  \Delta_1+2 t_c \omega _c \left(\gamma  m^4 \epsilon  t_c^2 \omega _c^2+\gamma  m^3 \epsilon ^2 t_c^3 \omega _c\right), \quad\text{and} \quad\Delta_1=\left(m \omega _0^2 t_c^2+\gamma  t_c+m\right){}^2+\left(\epsilon ^2-\alpha ^2\right) t_c^4.\nonumber \nonumber
    \end{equation}
    
 \begin{figure*}
    \centering
\includegraphics[width=0.75\linewidth]{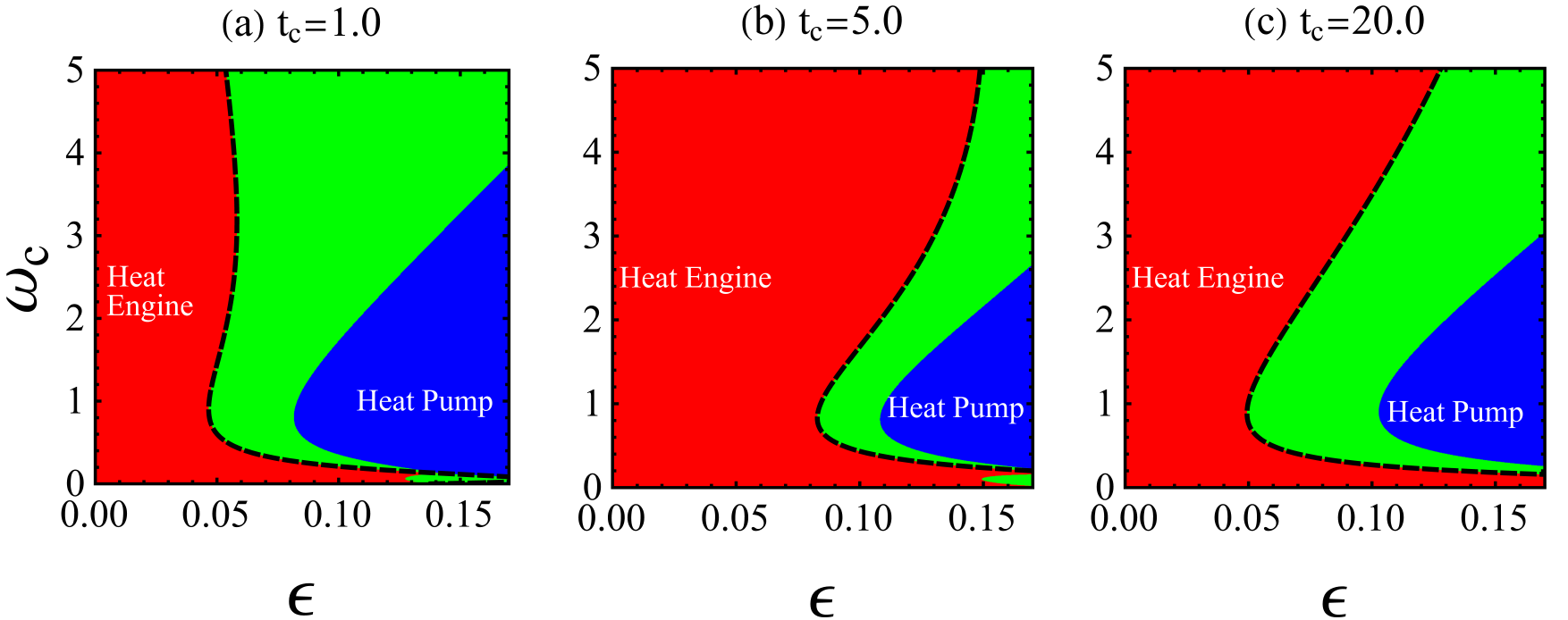}
    \caption{The regimes in $\epsilon$-$\omega_c$ parameter space separating the region for heat engine and heat pump in the case when the load is opposing the torque, is shown for $t_{c}=1.0$ in (a), $t_c=5.0$ in (b), and $t_c=20$ in (c), respectively. The other common parameters are $\alpha=0.75$, $\gamma=0.1$, $T_h=4.0$, $T_c=1.0$, and  $m=f_0=1$.}
    \label{fig:regimes}
\end{figure*}
In order to determine the efficiency, it is further required to evaluate the average rate at which heat is expelled from the hot bath ($Q_h$). Rearranging Eq.~\eqref{eq:Dynamics_OD}, we obtain
 \begin{equation}
    -\gamma  {\bf \dot{r}}(t)+{\bf f}^s(t)+{ {\boldsymbol \xi}}(t)= m \ddot{\bf r} - |q| \left( \dot{\bf r} \times {\bf B} \right) + \nabla U + {\bf F}_L.
     \label{eq:model_rearranged_OD}
  \end{equation}
Now, Using Eqs.~\eqref{eq:dQh}, ~\eqref{eq:Q_h_definition} and ~\eqref{eq:model_rearranged_OD}, the rate of heat from hot bath is given by
\begin{equation}
\begin{split}
    Q_h&=m \omega_c\langle\dot{x}\dot{y}\rangle+ (\alpha + \epsilon)\langle \dot{y}x \rangle
    \\
    &=-\frac{2 m^3 \omega_c}{S}\Biggr[2 \alpha  f_0^2 m t_c \omega _c \Bigr[m \left(-\gamma  \omega _0^2 t_c^2 \omega _c+\omega _0^2 \epsilon  t_c^2+\epsilon \right)+2 \gamma  \epsilon  t_c\Bigr]-2 \alpha  \gamma  f_0^2 t_c^3 \left(\alpha ^2+\gamma ^2 \omega _0^2-\epsilon ^2\right)\\
    &+\Delta _0 \Bigl(\alpha  \epsilon  \left(T_c+T_h\right)+\left(T_c-T_h\right) \left(-\gamma ^2 \omega _0^2-\gamma  \epsilon  \omega _c+\epsilon ^2\right)\Bigl)\Biggr]
    + \biggl(\frac{\alpha+\epsilon}{2S\epsilon}\biggl)\Biggl\{\Biggr[ 2 \gamma ^2 f_0^2 \epsilon ^2 t_c \bigl(m^2 \omega _0^2 t_c^2+\left(\gamma  t_c+m\right)^2 \bigl)\\
    &+2 f_0^2 m^2 \epsilon  t_c \omega _c\biggl( t_c^2 \left(\alpha ^2 \gamma +\gamma ^3 \omega _0^2-\epsilon \omega _c \left(\gamma ^2+m^2 \omega _0^2\right)+\gamma  m^2 \omega _0^2 \omega _c^2\right)-m^2 \left(\epsilon \omega _c-2 \gamma  m \epsilon  t_c \omega _c\right)\biggl)\Biggr]\\
    &+ \dfrac{\gamma   \epsilon\Delta_0}{m}  \left(\epsilon  \left(T_c+T_h\right) \left(\gamma ^2+m^2 \omega _c^2\right)+\alpha  \gamma ^2 \left(T_c-T_h\right)\right)\Biggl\}.
\end{split}   
\label{eq:Qh_OD_main}
\end{equation}
Similarly, the average rate of heat out of the cold bath $Q_c$ can be calculated using Eqs.~\eqref{eq:dQh}, ~\eqref{eq:Q_c_definition} and ~\eqref{eq:model_rearranged_OD} as
\begin{equation}
\begin{split}
    Q_c&=-m \omega_c\langle\dot{x}\dot{y}\rangle + (\alpha - \epsilon)\langle \dot{x}y \rangle
    \\
    &=\frac{2 m^3 \omega_c}{S}\Biggr[2 \alpha  f_0^2 m t_c \omega _c \Bigr[m \left(-\gamma  \omega _0^2 t_c^2 \omega _c+\omega _0^2 \epsilon  t_c^2+\epsilon \right)+2 \gamma  \epsilon  t_c\Bigr]-2 \alpha  \gamma  f_0^2 t_c^3 \left(\alpha ^2+\gamma ^2 \omega _0^2-\epsilon ^2\right)\\
    &+\Delta _0 \Bigl(\alpha  \epsilon  \left(T_c+T_h\right)+\left(T_c-T_h\right) \left(-\gamma ^2 \omega _0^2-\gamma  \epsilon  \omega _c+\epsilon ^2\right)\Bigl)\Biggr]
    + \biggl(\frac{\epsilon-\alpha}{2 S\epsilon}\biggl)\Biggl\{\Biggr[ 2 \gamma ^2 f_0^2 \epsilon ^2 t_c \bigl(m^2 \omega _0^2 t_c^2+\left(\gamma  t_c+m\right)^2 \bigl)\\
    &+2 f_0^2 m^2 \epsilon  t_c \omega _c\biggl( t_c^2 \left(\alpha ^2 \gamma +\gamma ^3 \omega _0^2-\epsilon \omega _c \left(\gamma ^2+m^2 \omega _0^2\right)+\gamma  m^2 \omega _0^2 \omega _c^2\right)-m^2 \left(\epsilon \omega _c-2 \gamma  m \epsilon  t_c \omega _c\right)\biggl)\Biggr]\\
    &+ \dfrac{\gamma   \epsilon\Delta_0}{m}    \left(\epsilon  \left(T_c+T_h\right) \left(\gamma ^2+m^2 \omega _c^2\right)+\alpha  \gamma ^2 \left(T_c-T_h\right)\right)\Biggl\}.
\end{split}   
\label{eq:Qc_OD_main}
\end{equation}
\end{widetext}

From Eq.~\eqref{eq:Qh_OD_main}, we observe that the total heat flow depends on the steady-state averages of the velocity cross-correlation and the position-velocity cross-correlation functions. The first term corresponds to the divergent heat flow in the system which vanishes in the $\omega_c\to0$ limit.


The condition $P > 0$ implies that the work can be extracted from the system. Since the load is in a direction opposite to that of the torque, condition $P > 0$ is necessary for the system to work as a heat engine. This is because the function of a heat engine is to deliver work using the expelled heat from the hot bath. The system while working against the load constantly dissipates heat by expelling it from the hot bath to the cold bath, 
which implies that $Q_h > 0$ for a heat engine. This further suggests that the necessary and sufficient condition for the system to operate as a heat engine is $Q_h > 0$ and $P > 0$. For such conditions, using Eq.~\eqref{eq:Eff_main} the efficiency $\eta$ of the heat engine can be calculated as
\begin{equation}
    \eta=\dfrac{2 \epsilon }{\alpha +\epsilon+ 2 \lambda},
    \label{eq:Eff_1st}
\end{equation}
where 
$\lambda$ is given by
\begin{widetext}
    \begin{equation}
        \lambda=\dfrac{\omega _c \left[2 \alpha  f_0^2 m \epsilon  t_c \left(m \omega _0^2 t_c^2+2 \gamma  t_c+m\right)+\gamma \Delta _0 \left(\alpha  \epsilon  \left(T_c+T_h\right)+\left(T_c-T_h\right) \left(\epsilon ^2-\gamma ^2 \omega _0^2\right)\right)+m \omega _0^2 t_c^2 \left(\epsilon -\gamma \omega _c\right)+2 \gamma  \epsilon  t_c+m \epsilon \right]}{\omega _c \left[f_0^2 m^2 t_c \left(t_c \left(2 \gamma  m \epsilon  \omega _c-t_c \left(\gamma  \left(\alpha ^2-\gamma  \epsilon  \omega _c\right)+\omega _0^2 \left(\gamma ^3+m^2 \omega _c \left(\gamma  \omega _c-\epsilon \right)\right)\right)\right)+m^2 \epsilon  \omega _c\right)+\gamma  \Delta _0 m^2 \epsilon  \omega _c \left(T_c-T_h\right)\right]+S_3},
        \label{eq:lambda}
    \end{equation}
with $S_3$ as
    \begin{equation}
        S_3=\gamma ^2 \left[f_0^2 \epsilon  t_c \left(m^2 \omega _0^2 t_c^2+\left(\gamma  t_c+m\right)^2 \right)+\gamma  \Delta _0  (\alpha +\epsilon ) \left(T_c-T_h\right)\right].
    \end{equation}        
\end{widetext}

\begin{figure}
    \centering
    \includegraphics[width=\linewidth]{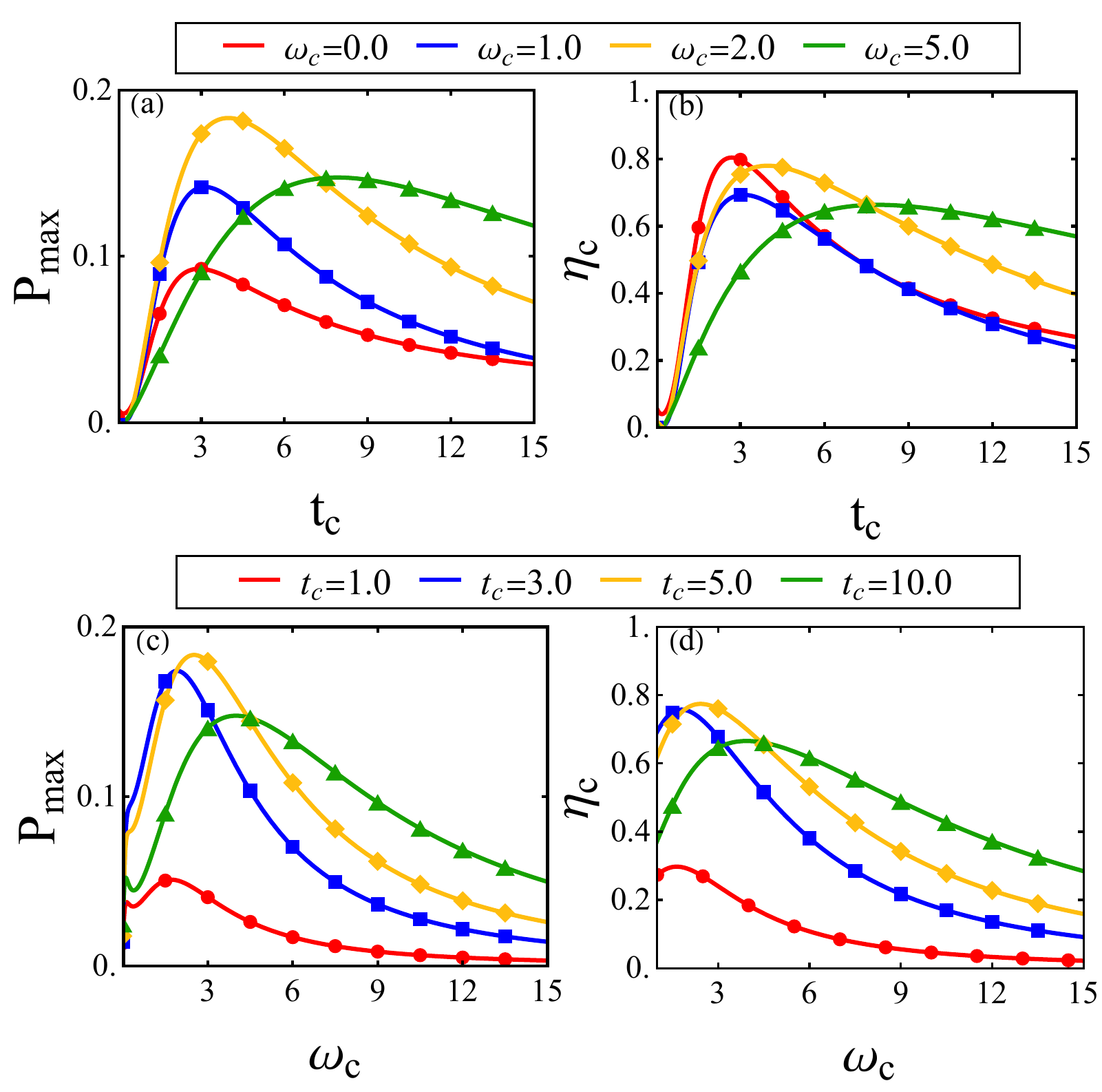}
    \caption{$P_{\text{max}}$ and $\eta_{c}$ as a function of $t_c$ for different values of $\omega_c$ is shown in (a) and (b), respectively. $P_{\text{max}}$ and $\eta_c$ as a function of $\omega_c$ for different $t_c$ values is shown in (c) and (d), respectively. The other common parameters are $m = f_0 = \omega_0 = 1.0$, $\gamma=0.1$, $\alpha=0.75$, $T_h=4.0$, and $T_c=1.0$.}
    \label{fig:Pmax_eta}
\end{figure}
\begin{figure}
\centering
\includegraphics[width=\linewidth]{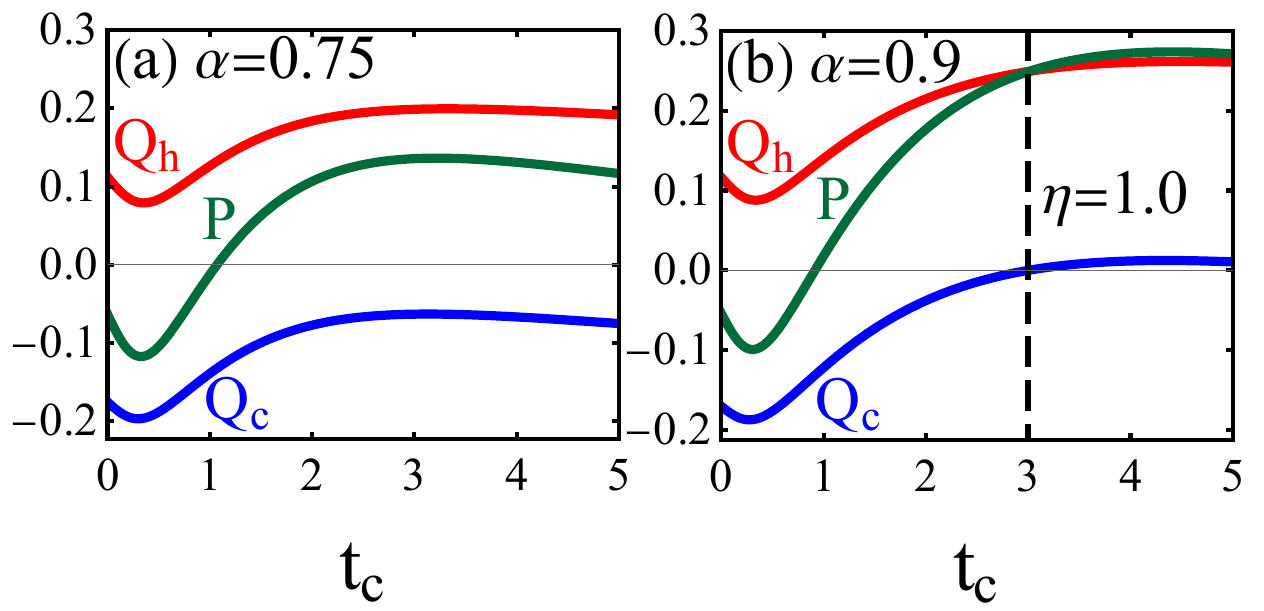}
    \caption{The evolution of $P$, $Q_h$ and $Q_c$ as a function of  $t_c$ is shown in (a) for $\alpha=0.75$ and in (b) for $\alpha=0.9$. Other common parameters are $\epsilon=0.05$, $\gamma=0.1$, $T_c=1.0$, $T_h=4.0$, and $m = f_0 = \omega_0 = \omega_c =1.0$.}
    \label{fig:p_Qh_Qc}
\end{figure}

Similarly, the system can also operate as a heat pump by reversing the direction of heat flow. This can be achieved by loading the engine beyond the stall until the heat flow changes direction so that the heat is absorbed by the hot bath, i.e., $Q_h<0$. This condition implies that heat flow is against the temperature gradient. It is possible only when work is done on the system, i.e., $P<0$. Therefore, the system can operate as a heat pump only when both $P$ and $Q_h$ are negative. In such a situation, the performance characteristics of the heat pump can be determined by calculating the COP using Eq.~\eqref{eq:COP_main}, and it is given by
\begin{equation}
    \text{COP}=\frac{\lambda}{\epsilon}+\frac{\alpha}{2\epsilon}+\frac{1}{2}.
    \label{eq:COP_1st}
\end{equation}

In Fig.~\ref{fig:regimes}, based on the calculation of $P$ and $Q_h$, we have shown the regimes in which the system works as a heat engine or heat pump in $\omega_c - \epsilon$ parameter space and for three different values of $t_c$. When the applied load strength $\epsilon$ is very small, both $P$ and $Q_h$ are positive (red region in Fig.~\ref{fig:regimes}), and the system performs as a heat engine. As $\epsilon$ increases by keeping the value of $\omega_c$ fixed, at a specific value of $\epsilon$ ($=\epsilon_s$), the power delivered by the system becomes zero ($P = 0$), causing the engine to stall. This stalling parameter $\epsilon_s$ corresponding to each $\omega_c$ values defines the boundary of the red region in the $\omega_c - \epsilon$ parameter space, shown as a black dashed line in Fig.~\ref{fig:regimes}, along which $P=0$. Beyond this $\epsilon_s$, if the system is further loaded, the power becomes negative ($P<0$) while $Q_h$ still remains positive ($Q_h > 0$), i.e., the green region in Fig.~\ref{fig:regimes}, implying that the system in this regime with $P<0$ and $Q_h >0$, can function neither as a heat engine nor as a heat pump. With further increase in $\epsilon$ value, at $\epsilon= \epsilon_r$, $Q_h$ vanishes ($Q_h = 0$). Beyond this $\epsilon_r$, both $P$ and $Q_h$ become negative and the system performs as a heat pump  (blue region). Moreover, in $\omega_c \to 0$ limit, for higher values of $\gamma$, where the inertial effects are negligible, $\epsilon_s$ becomes same as $\epsilon_r$. This indicates a direct transformation of the system from the parameter regime of the heat engine to the heat pump regime as $\epsilon$ increases. 
From Fig.~\ref{fig:regimes}, It is observed that with an increase in the $t_c$ value, the parameter regime in $\omega_c-\epsilon$ space, for which it works as a heat engine enhances [Figs.~\ref{fig:regimes}(a) and (b)], it finally reaches at a maximum spread, and then further gets reduced for higher values of $t_c$ [Fig.~\ref{fig:regimes}(c)]. However, the parameter regime for which the system works as a refrigerator decreases, reaches a minimum spread and then increases with further increase in $t_c$ values. 

By tuning the strength of load force $\epsilon$, one can extract the maximum power ($P_{\text{max}}$) delivered by the system using 
\begin{equation}
    \frac{\partial P}{\partial \epsilon}\Bigg|_{\epsilon = \epsilon_{\text{max}}} = 0.
    \label{eq:Pmax_def}
\end{equation}
We have plotted $P_{\text{max}}$ and the efficiency $\eta_c$ corresponding to $P_{\text{max}}$ as a function of $t_c$ for different values of $\omega_c$ in Figs.~\ref{fig:Pmax_eta}(a) and (b), respectively.
Both $P_{\text{max}}$ and $\eta_c$ show a non-monotonic behavior with $t_c$. Initially, $P_{\text{max}}$ and $\eta_c$ increase with $t_c$, attain a maximum value for an intermediate $t_c$ value, and then decrease with a further increase in $t_c$ values. As the value of $\omega_c$ increases, the maximum value of $P_{\text{max}}$ initially increases and moves toward higher values of $t_c$. Furthermore, for higher values of $\omega_c$, the peak of $P_{\text{max}}$ is suppressed, and the curve of $P_{\text{max}}$ becomes broader. However, the maximum value of $\eta_c$ shows a non-monotonic dependence on $\omega_c$. 
Similarly, in Figs.~\ref{fig:Pmax_eta}(c) and (d), we have plotted $P_{\text{max}}$ and $\eta_c$ as functions of $\omega_c$ for different values of $t_c$. The behavior of both $P_{\text{max}}$ and $\eta_c$ with $\omega_c$ is non-monotonic, similar to the case discussed above.
\begin{figure*}[!ht]
    \centering
    \includegraphics[width=0.7\linewidth]{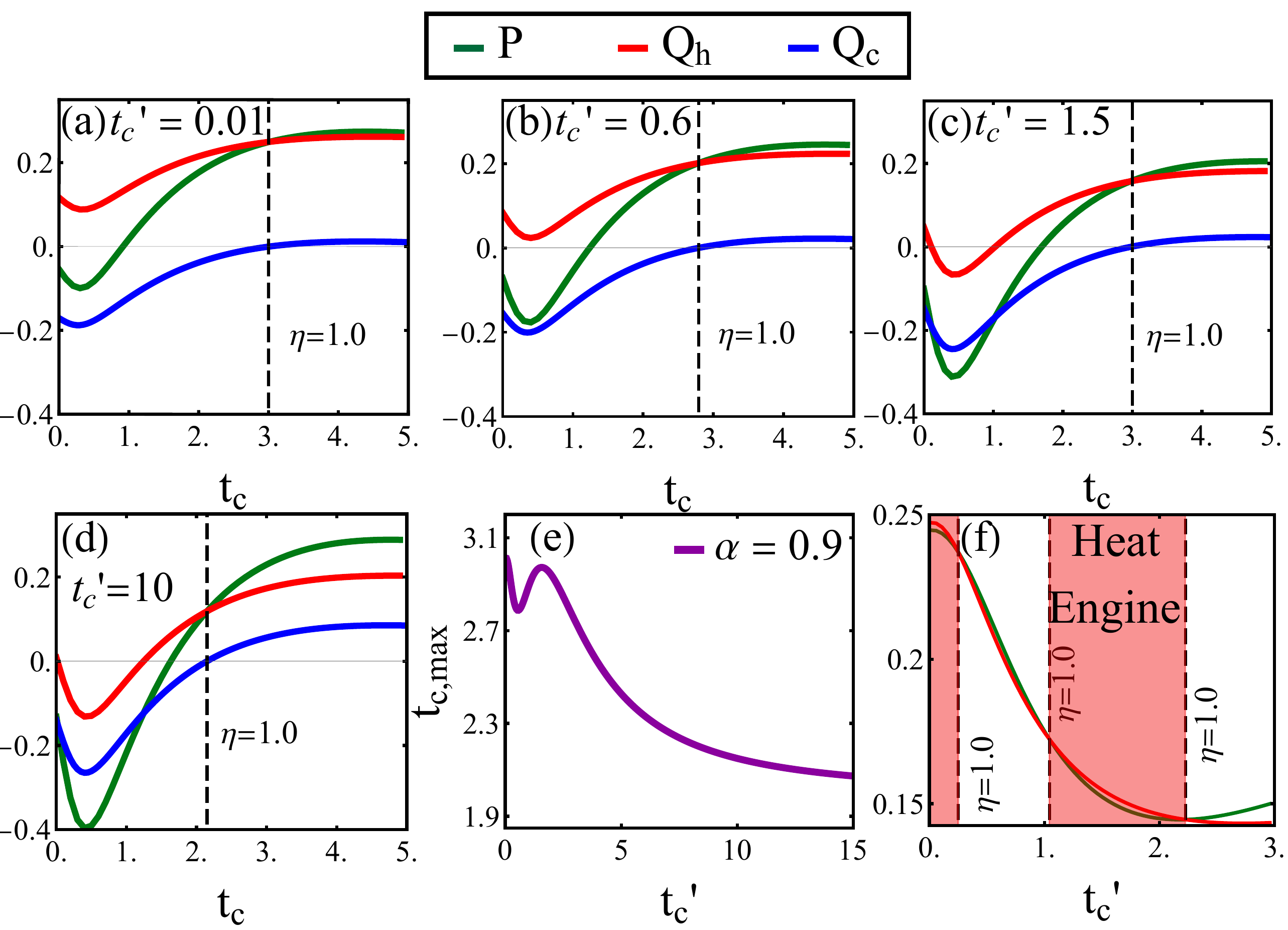}
    \caption{The evolution of $P$, $Q_h$ and $Q_c$ as a function of  $t_c$ is plotted in (a) for $t_c'=0.01$, in (b) for $t_c'=0.6$, in (c) for $t_c'=1.5$ and in (d) for $t_c'=10$. (e) shows the evolution of $t_{c,\text{max}}$ as function of $t_c'$ for a fixed $\alpha=0.9$. (f) The evolution of $P$ and $Q_h$ as function of $t_c'$ for $t_c=2.9$, corresponding to the plot in (e). The other common parameters are $\epsilon=0.05$, $\gamma=0.1$, $T_c=1.0$, $T_h=4.0$, and $m = f_0 = \omega_0 = \omega_c =1.0$. }
    \label{fig:viscoelastic_OD}
\end{figure*}
Initially, when the value of $\omega_c$ increases, both $P_{\text{max}}$ and $\eta_c$ also increase, reach a maximum value, and then decrease with a further increase in the value of $\omega_c$. The peak value of $P_{\text{max}}$ and $\eta_c$ also shows a similar non-monotonic dependence with $t_c$.

In Fig.~\ref{fig:p_Qh_Qc}, we have plotted $Q_h$, $Q_c$, and $P$ as a function of $t_c$ for two different values of $\alpha$. For smaller values of $t_c$, $P$ is negative while $Q_h$ remains positive, indicating that the system does not perform as a heat engine. As $t_c$ increases, there exists a critical value of $t_c$, at which $Q_h = -Q_c$, leading to $P = 0$. Beyond this point, $P$ becomes positive and the system starts to function as a heat engine.
For small values of $\alpha$ [Fig.~\ref{fig:p_Qh_Qc}(a)], $Q_h$ is always positive, and $Q_c$ is always negative, ensuring that $P$ is always less than $Q_h$, and consequently the efficiency of the system always remains less than one. However, for a larger value of $\alpha$ [Fig.~\ref{fig:p_Qh_Qc}(b)], $Q_c$ can have zero value for a finite $t_c$ ($=t_{c,\text{max}}$). Under this condition, $P$ becomes the same as $Q_h$. This implies that all heat from the hot bath is converted into mechanical work without transferring any heat to the low temperature (cold) bath, which leads to the efficiency of the system being $100\%$. Thus, the efficiency of an active magneto-gyrator has no upper bound, and can be brought close to one by tuning the system parameters.  

Further, when the system is suspended in a viscoelastic bath characterized by a time-dependent memory kernel, the Eq.~\eqref{eq:Dynamics_OD} becomes
\begin{widetext}
\begin{equation}
    \ddot{{\bf r}}(t) = -\int\limits_0^{t}f(t-t'){\bf \dot{r}}(t')\;dt' - \omega_c {\bf \dot{r}}(t)\times\hat{k}
    -\dfrac{\nabla U}{m} + \frac{{\bf f}^s(t)}{m} - \frac{{\bf F}_L}{m}+ \frac{{{\boldsymbol{\eta}}}(t)}{m}.
\end{equation}
\end{widetext}
Here, the memory kernel $f(t-t')$ has the form
\begin{equation}
    f(t-t') = \frac{\gamma}{2}\delta(t-t') + \frac{\gamma}{2 t_c'} e^{-(t-t')/t_c'},
    \label{eq:memory_kernel}
\end{equation}
with the parameter $t_c'$ representing the viscoelastic memory timescale. It represents the timescale up to which there persists a transient elasticity in the medium. Noise $\boldsymbol{\eta}(t)$ has statistical properties $\langle \boldsymbol{\eta}(t) \rangle = 0$ and satisfies the generalized fluctuation dissipation relation $\langle \boldsymbol{\eta}_i(t) \boldsymbol{\eta}_j(t') \rangle = \delta_{ij}T_i f(t-t')$.
Similar to the viscous limit, when the load is applied in the opposite direction, the system functions as a heat engine. In Figs.~\ref{fig:viscoelastic_OD}(a)-(d), we have plotted $P$, $Q_h$, and $Q_c$ as functions of $t_c$ for different values of $t_c'$. For a fixed $t_{c}'$, as the value of $t_c$ increases, the value of $\eta$ increases and reaches $\eta = 1$ at a finite value of $t_c$($ = t_{c,\text{max}}$) [see Figs.~\ref{fig:viscoelastic_OD}(a)-(d)]. As $t_c'$ increases, $t_{c,\text{max}}$ initially decreases from a higher value, then increases with an increase in $t_c'$, and finally decreases again, eventually saturating to a constant value for a larger value of $t_c'$. 
This non-monotonic dependence of $t_{c,\text{max}}$ on $t_c'$ is shown in Fig.~\ref{fig:viscoelastic_OD}(e). This is further explored in Fig.~\ref{fig:viscoelastic_OD}(f), where we have plotted $P$ and $Q_h$ as a function of $t_c'$ for a fixed value of $t_{c,\text{max}} = 2.9$. The shaded region represents the values of $t_c'$ where the system functions as a heat engine. 
These are the only regions where the efficiency can be clearly defined and is less than one since $P<Q_h$. It is to be noted that for the parameters at the boundary of these regions (i.e., the $t_c'$ value corresponding to the dashed line), $\eta$ becomes unity, which implies that for a fixed $t_{c,\text{max}}$, there exists more than one value of $t_c'$ at which the efficiency becomes $100\%$.

\begin{figure}
    \centering
    \includegraphics[width=\linewidth]{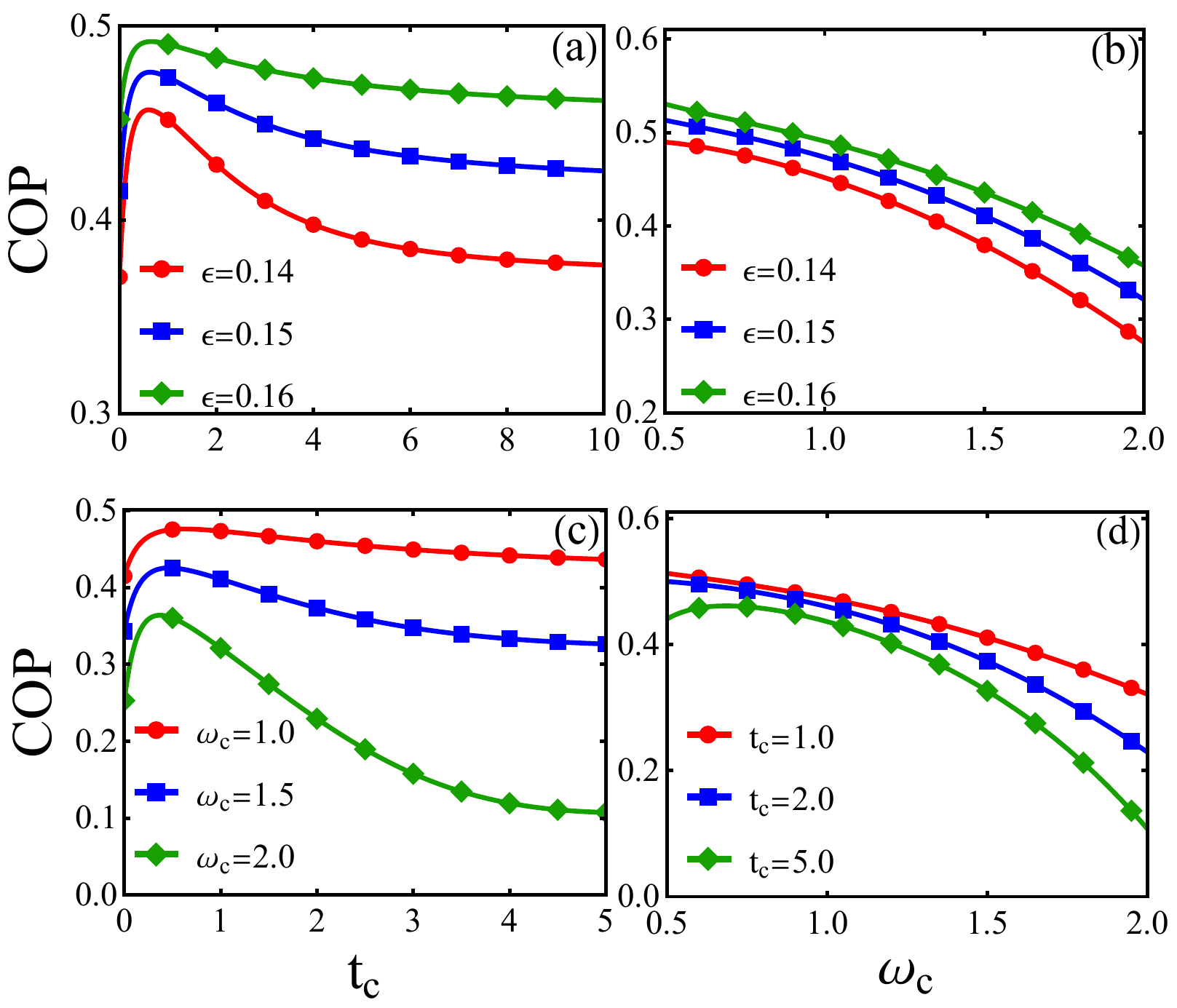}
    \caption{The COP as a function of $t_c$ is shown in (a) for different values of $\epsilon$ and in (c) for different values of  $\omega_c$. Similarly, the COP as a function of $\omega_c$ is shown in (b) for different values of $\epsilon$ and in (d) for different values of $t_{c}$. In (c) and (d), $\epsilon=0.15$ is taken. The other common parameters are  $m = f_0 = 1.0$, $\gamma=0.1$, $\alpha=0.75$, $T_c=1.0$, $T_h=4.0$, and $\omega_0=1.0$.}
    \label{fig:COP_tc_wc_OD}
\end{figure}

The regimes where the system functions as a heat pump are shown in blue in Fig.~\ref{fig:regimes}. In order to understand the performance of heat pump in these parameter regimes, in Fig.~\ref{fig:COP_tc_wc_OD}, we have plotted the COP as a function of $t_c$ in Fig.~\ref{fig:COP_tc_wc_OD} (a) and in Fig.~\ref{fig:COP_tc_wc_OD} (b) for different values of $\epsilon$ and $\omega_c$, respectively. The COP exhibits a non-monotonic dependence on $t_c$ [Figs.~\ref{fig:COP_tc_wc_OD}(a) and (c)]. It increases, reaches a maximum value for a finite $t_c$, and then decreases for higher values of $t_c$. Additionally, increasing $\epsilon$ value enhances the COP [Fig.~\ref{fig:COP_tc_wc_OD}(a)], whereas increasing the $\omega_c$ reduces it [Figs.~\ref{fig:COP_tc_wc_OD}(c)]. The behavior of COP with $\omega_c$ is monotonic, as it decreases with an increase in $\omega_c$ value [Figs.~\ref{fig:COP_tc_wc_OD}(b) and (d)]. However, an increase in $\epsilon$ results in an increase in COP. 

In the limit $t_c \to 0$ or in the vanishing limit of activity timescale, the system behaves as a passive magneto gyrator, and the expression of $P$ [Eq.~\eqref{eq:Power_FE_OD}] reduces to
\begin{equation}
 P= \frac{\gamma   \epsilon  \left(\epsilon  \left(T_c+T_h\right) \left(\gamma ^2+m^2 \omega _c^2\right)+\alpha  \gamma ^2 \left(T_c-T_h\right)\right)}{m\left[\left(\gamma ^2+m^2 \omega _c^2\right) \left(-\gamma ^2 \omega_0^2-\gamma  \epsilon  \omega _c+\epsilon ^2\right)-\alpha ^2 \gamma ^2\right]}.
\label{eq:P_tc0}
 \end{equation}
Similarly, the expression for $Q_h$ in this limit ($t_c \to 0$) is given by
\begin{equation}
   \begin{split}     Q_h=&\dfrac{\gamma\epsilon  \left(T_c+T_h\right) \left(\gamma ^2 \epsilon +m \omega _c^2 (m (\alpha +\epsilon )-\alpha )\right)}{2 m \left[\left(-\gamma ^2 \omega _0^2-\gamma  \epsilon  \omega _c+\epsilon ^2\right) \left(\gamma ^2+m^2 \omega _c^2\right)-\alpha ^2 \gamma ^2\right]}\\
   &+\dfrac{\gamma   \left(T_c-T_h\right) \left(\alpha ^2 \gamma ^2+m \omega _c^2 \left(\gamma ^2 \omega _0^2+\gamma  \epsilon  \omega _c-\epsilon ^2\right)\right)}{2 m\left[\left(-\gamma ^2 \omega _0^2-\gamma  \epsilon  \omega _c+\epsilon ^2\right) \left(\gamma ^2+m^2 \omega _c^2\right)-\alpha ^2 \gamma ^2\right]}.
   \end{split}
   \label{eq:Qh_special}
\end{equation}
\begin{figure}
\includegraphics[width=0.75\linewidth]{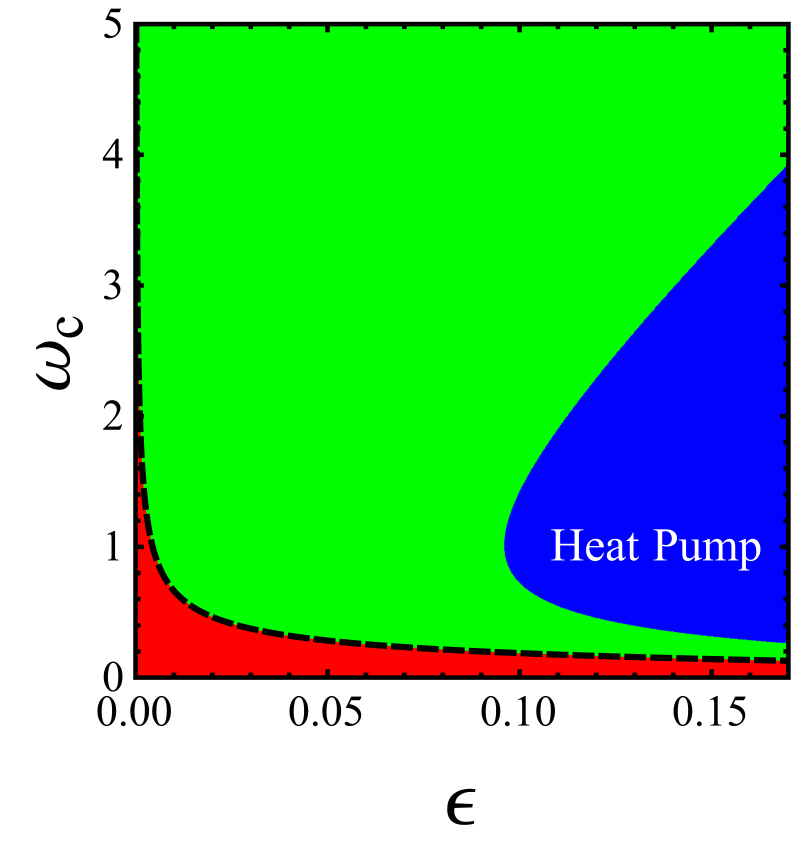}
\caption{The parameter regimes (in $t_c \to 0$ limit) in $\epsilon-\omega_c$ parameter space, where the system works as a heat engine or heat pump. The common parameters are $m = f_0 = 1.0$, $\gamma=0.1$, $\alpha=0.75$, $\omega_0=1.0$, $T_h=4$, and $T_c=1$.}
    \label{fig:tc0_regime_OD}
\end{figure}
The system in this limit can function as a heat engine or a heat pump by varying the load strength, as shown in Fig.~\ref{fig:tc0_regime_OD}. However, the parameter regime where the system operates as a heat engine (red region in Fig.~\ref{fig:tc0_regime_OD}) is significantly smaller compared to the red regimes in the active magneto-gyrator case [see Fig.~\ref{fig:regimes}]. In this limit, the maximum power and efficiency can be calculated from Eqs.~\eqref{eq:Pmax_def}-\eqref{eq:Qh_special} and Eq.~\eqref{eq:Eff_main}, respectively. 
\begin{figure}
     \centering
     \includegraphics[width=\linewidth]{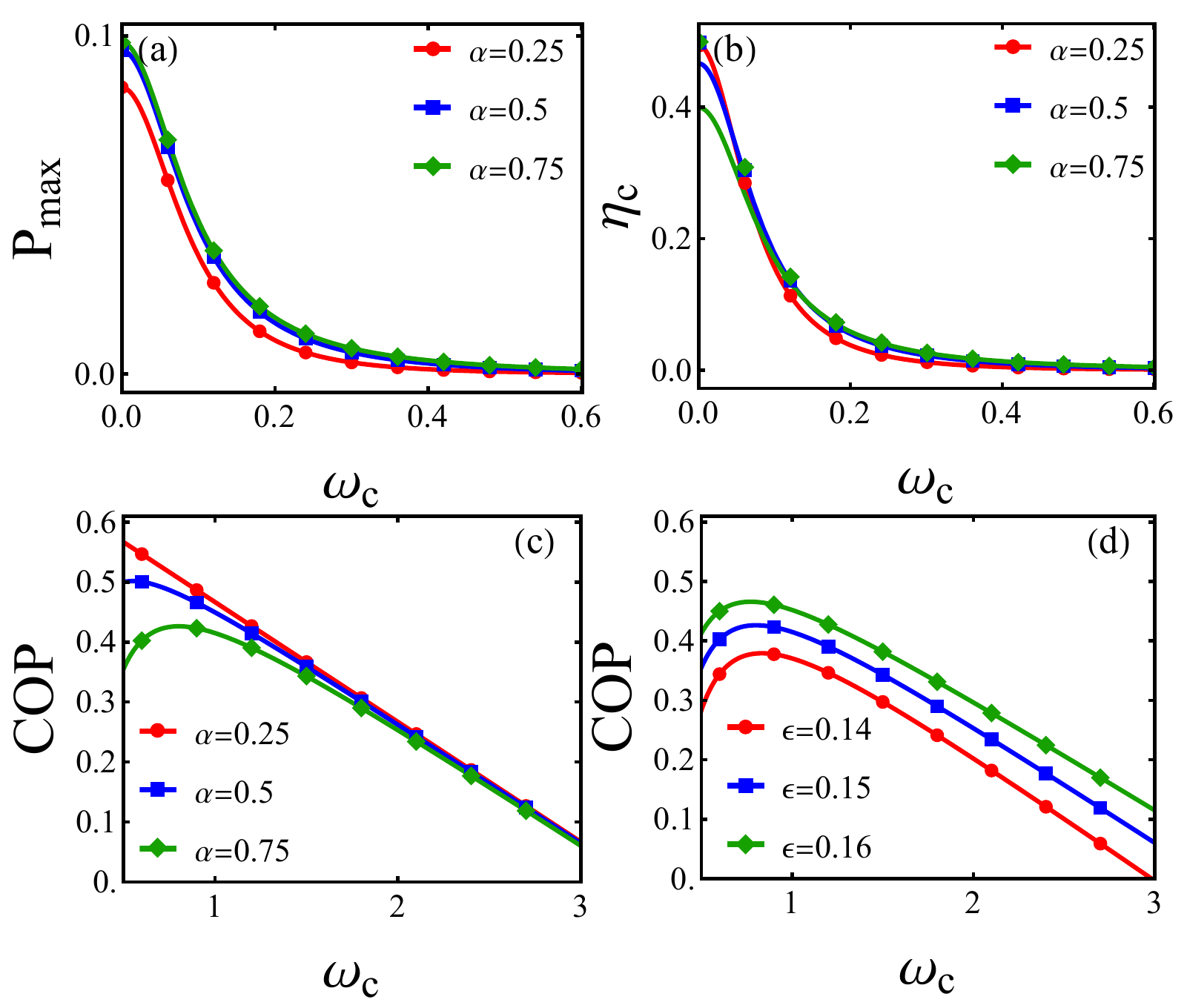}
     \caption{$P_{\text{max}}$ and $\eta$ as a function of $\omega_c$ for different values of $\alpha$ and $\epsilon$ is shown in (a) and (b), respectively. The COP as a function of $\omega_c$ for different values of $\alpha$ and $\epsilon$ is shown in (c) and (d), respectively. The common parameters are  $m = f_0 = 1.0$, $\gamma=0.1$, and $\omega_0=1.0$.}
     \label{fig:tc0}
\end{figure}
In Figs.~\ref{fig:tc0}(a) and (b), we have plotted $P_\text{max}$ and the corresponding efficiency $\eta$ of the heat engine as a function of $\omega_c$ for different values of $\alpha$ and $\epsilon$, respectively.
We observe that both $P_\text{max}$ and $\eta$ decrease with increase in the value of $\omega_c$. Furthermore, while $P_\text{max}$ increases with $\alpha$ [Fig.~\ref{fig:tc0}(a)], $\eta$ shows a minimal change with $\alpha$ in the absence of activity [Figs.~\ref{fig:tc0}(b)].
Similarly, in Figs.~\ref{fig:tc0}(c) and (d), we have shown the COP plot of the heat pump (in the heat pump regime) as a function of $\omega_c$ for different values of $\alpha$ and $\epsilon$, respectively. In low $\omega_c$ regime, the COP exhibits a non-monotonic dependence on $\omega_c$ for smaller $\alpha$ values [Fig.~\ref{fig:tc0}(c)]. For higher $\omega_c$ values, the COP decreases linearly with $\omega_{c}$. Increase of $\epsilon$ results increase in the COP [Fig.~\ref{fig:tc0}(d)].

In the limit $\omega_c \to 0$, the dynamics [Eq.~\eqref{eq:Dynamics_OD}] reduces to the case of an active Brownian gyrator. In this case, the power $P$ in Eq.~\eqref{eq:Power_FE_OD} is given by
\begin{equation}  
\begin{split}
P&=\dfrac{2 f_0^2 \epsilon ^2 t_c \bigl(m^2 \omega _0^2 t_c^2+\left(\gamma  t_c+m\right)^2 \bigl)}{m \Delta_1  \left(\alpha^2+\gamma^2 
\omega_0^2-\epsilon^2\right)}\\
&+\dfrac{\gamma\epsilon\left[\alpha  \left(T_h-T_c\right)+\epsilon  \left(T_c+T_h\right)\right]}{m  \left(\alpha ^2+\gamma ^2 \omega _0^2-\epsilon ^2\right)}.
\label{eq:Power_FE_wmf}
\end{split}
\end{equation}
Thus, the power can be extracted from the system even in the absence of an applied magnetic field. The corresponding values of $Q_h$ [Eq.~\eqref{eq:Qh_OD_main}] and $Q_c$ [Eq.~\eqref{eq:Qc_OD_main}] in this limit are given by
\begin{equation}  
\begin{split}
Q_h&=\dfrac{\epsilon+\alpha}{2 \epsilon}\Biggl\{\dfrac{2 f_0^2 \epsilon ^2 t_c \bigl(m^2 \omega _0^2 t_c^2+\left(\gamma  t_c+m\right)^2 \bigl)}{m \Delta_1  \left(\alpha^2+\gamma^2 
\omega_0^2-\epsilon^2\right)}\\
&+\dfrac{\gamma\epsilon\left[\alpha  \left(T_h-T_c\right)+\epsilon  \left(T_c+T_h\right)\right]}{m  \left(\alpha ^2+\gamma ^2 \omega _0^2-\epsilon ^2\right)}\Biggl\},
\label{eq:Q_h_wc0}
\end{split}
\end{equation}
and
\begin{equation}  
\begin{split}
Q_c&=\dfrac{\epsilon-\alpha}{2 \epsilon}\Biggl\{\dfrac{2 f_0^2 \epsilon ^2 t_c \bigl(m^2 \omega _0^2 t_c^2+\left(\gamma  t_c+m\right)^2 \bigl)}{m \Delta_1  \left(\alpha^2+\gamma^2 
\omega_0^2-\epsilon^2\right)}\\
&+\dfrac{\gamma\epsilon\left[\alpha  \left(T_h-T_c\right)+\epsilon  \left(T_c+T_h\right)\right]}{m  \left(\alpha ^2+\gamma ^2 \omega _0^2-\epsilon ^2\right)}\Biggl\}.
\label{eq:Qh_Qc_wc0}
\end{split}
\end{equation}

In this limit ($\omega_c \to 0$), the value of $\lambda$ in Eq.~\eqref{eq:lambda} vanishes and hence, the efficiency [Eq.~\eqref{eq:Eff_1st}] of an active Brownian gyrator $\eta_{AG}$ reduces to
\begin{equation}
    \eta_{AG}=\dfrac{2 \epsilon }{\alpha +\epsilon}.
    \label{eq:eta_AG}
\end{equation}
When $\epsilon = \alpha$, the value of $Q_c$ vanishes as evident from Eq.~\eqref{eq:Qh_Qc_wc0}, and consequently $P$ becomes equal to $Q_h$, making $\eta_{AG}$ to be unity. But in this regime ($\omega_c \to 0$ limit) with $\alpha=\epsilon$, both $P$ and $Q_h$ become negative, as clearly evident from Eqs.~\eqref{eq:Power_FE_wmf} and ~\eqref{eq:Q_h_wc0}. Thus, the system does not operate as a heat engine in this limit. It should be noted that the Eq.~\eqref{eq:eta_AG} is valid only for a heat engine and hence is not applicable when the corresponding $P$ and $Q_h$ are in the heat pump or unstable regimes. 

\subsection{Applying Load in the same direction of torque }
\begin{figure*}
\centering
\includegraphics[width=0.75\linewidth]{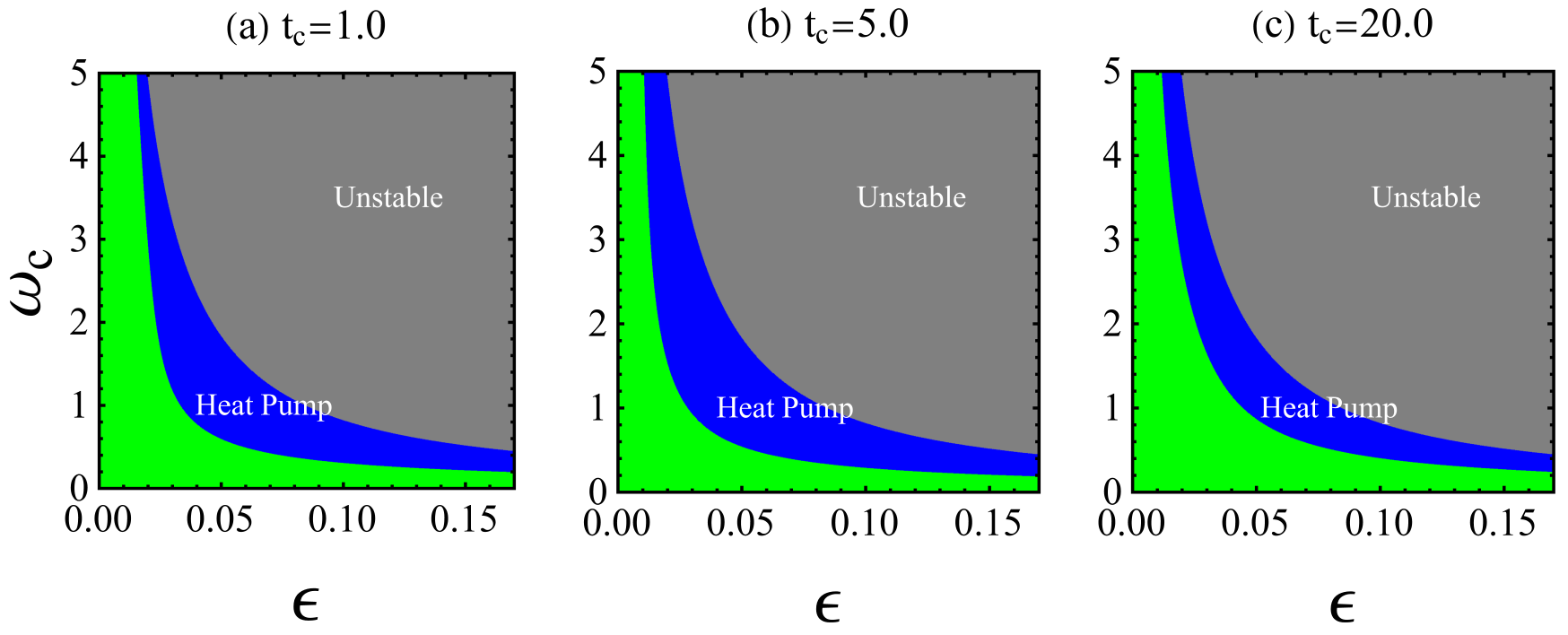}
\caption{Parameter regimes where the system works as a heat engine or a heat pump are shown in $\epsilon-\omega_c$ parameter space, for three different $t_c$ values in (a), (b), and (c), respectively.  The other common parameters are $m = f_0 = 1.0$, $\gamma=0.1$, $\alpha=0.75$, $T_h=4.0$, $T_c=1.0$ and $\omega_0=1.0$.}
\label{fig:Regimes_sd}
\end{figure*}
When the load ${\bf F}_L$ acts in the same direction of the torque, the Eq.~\eqref{eq:maineqmotion-vector} can be rewritten as
\begin{equation}
     \ddot{{\bf r}}(t) = -\frac{\gamma}{m} \dot{{\bf r}}(t)-\omega_c {\bf \dot{r}}(t)\times\hat{k}-\dfrac{\nabla U}{m} + \frac{{\bf f}^s(t)}{m} + \frac{{\bf F}_L}{m}+ \frac{{{\boldsymbol{\xi}}}(t)}{m},
     \label{eq:Dynamics_SD}
\end{equation}

From Fig.~\ref{fig:torque_DR}, it is already confirmed that the torque is always positive for positive $\alpha$ values. Hence, by applying a load in the same direction as that of the effective torque, it is possible to do work on the system. However, the work cannot be extracted from the system. Thus, the mechanical power $P$ is always negative. This suggests that the system cannot function as a heat engine. However, by tuning the parameters, the system can function as a heat pump. 
In this case, seeking the dynamics Eq.~\eqref{eq:Dynamics_SD}, the Eq.~\eqref{eq:dynamics-matrix} of section \ref{sec:A} can be rewritten as 
\begin{equation}
    \dot{\boldsymbol{\chi}} = A'\boldsymbol{\chi} + B \boldsymbol{\xi'},
    \label{eq:dynamics-matrix2}
\end{equation}
where the matrix $A'$ takes the form
\begin{equation}
    A' = \begin{pmatrix}
        0 & 0 & 1 & 0 & 0 & 0 \\
        0 & 0 & 0 & 1 & 0 & 0 \\
        -\omega_0^2 & \frac{-\alpha-\epsilon}{m} & -\frac{\gamma}{m} & \omega_c & \frac{\sqrt{D}}{m} & 0 \\ 
        \frac{-\alpha+\epsilon}{m} & -\omega_0^2 & -\omega_c & -\frac{\gamma}{m}  & 0 & \frac{\sqrt{D}}{m} \\
        0 & 0 & 0 & 0 & \frac{-1}{t_c} & 0 \\
        0 & 0 & 0 & 0 & 0 & \frac{-1}{t_c}
    \end{pmatrix}.
\end{equation}
Following the same procedure as in section \ref{sec:A}, the correlation matrix ${\bf \Xi'}$ [See Eq.~\eqref{eq:corr_mat_def}] can satisfy the equation
\begin{equation}
    A'\cdot {\bf \Xi'} + {\bf \Xi'} \cdot {A'}^T + B\ B^T = 0.
    \label{eq:Xi_relation_case2}
\end{equation}
Now, by solving the above equation, one can obtain ${\bf \Xi'}$, with which all the relevant physical quantities can be calculated. Using Eq.~\eqref{eq:power_main}, the mechanical power $P$ can be calculated as
\begin{widetext}
    \begin{equation}  
    \begin{split}
P&=-\epsilon\langle x\dot{y}-y \dot{x}\rangle\\
&= \Biggr[-2 f_0^2 m^2 \epsilon  t_c \omega _c\biggl\{\dfrac{ \bigl[t_c^2 \left(\alpha ^2 \gamma +\gamma ^3 \omega _0^2+\epsilon \omega _c \left(\gamma ^2+m^2 \omega _0^2\right)+\gamma  m^2 \omega _0^2 \omega _c^2\right)-m^2 \left(-\epsilon \omega _c+2 \gamma  m \epsilon  t_c \omega _c\right)\bigl]}{S_2}\biggl\}\\
&-\dfrac{2 \gamma ^2 f_0^2 \epsilon ^2 t_c \bigl(m^2 \omega _0^2 t_c^2+\left(\gamma  t_c+m\right)^2 \bigl)}{S_2}\Biggr]- \dfrac{\gamma  \epsilon\Delta_2}{m}\left(\dfrac{-\epsilon  \left(T_c+T_h\right) \left(\gamma ^2+m^2 \omega _c^2\right)+\alpha  \gamma ^2 \left(T_c-T_h\right)}{S_2}\right),
\label{eq:Power_FE}
\end{split}
    \end{equation}
where
    \begin{equation}
      S_2=\Delta _2 \left(\alpha ^2 \gamma ^2+\left(\gamma ^2 \omega _0^2+\gamma  \epsilon  \omega _c-\epsilon ^2\right) \left(\gamma ^2+m^2 \omega _c^2\right)\right)
      \label{eq:S2_def}
    \end{equation}
    and
    \begin{equation} 
        \Delta_2=  m \left[\left(m \omega _0^2 t_c^2+\gamma  t_c+m\right){}^2+\left(\epsilon ^2-\alpha ^2\right) t_c^4\right]+2 t_c \omega _c \left(-\gamma  m^4 \epsilon  t_c^2 \omega _c^2+\gamma  m^3 \epsilon ^2 t_c^3 \omega _c\right).\nonumber
    \end{equation}
Similarly, using Eq.~\eqref{eq:dQh}, the average heat out of the hot bath $Q_h$ can be calculated as
\begin{equation}
\begin{split}
    Q_h&=m \omega_c\langle\dot{x}\dot{y}\rangle-\dfrac{\epsilon}{2}\langle x\dot{y}-y \dot{x}\rangle-\dfrac{\alpha}{2}\langle x\dot{y}-y \dot{x}\rangle
    \\
    &=-\frac{2 m^3 \omega_c}{S_2}\Biggr[2 \alpha  f_0^2 m t_c \omega _c \Bigr[m \left(-\gamma  \omega _0^2 t_c^2 \omega _c+\omega _0^2 \epsilon  t_c^2+\epsilon \right)+2 \gamma  \epsilon  t_c\Bigr]-2 \alpha  \gamma  f_0^2 t_c^3 \left(\alpha ^2+\gamma ^2 \omega _0^2-\epsilon ^2\right)\\
    &+\Delta _2 \Bigl(\alpha  \epsilon  \left(T_c+T_h\right)+\left(T_c-T_h\right) \left(-\gamma ^2 \omega _0^2-\gamma  \epsilon  \omega _c+\epsilon ^2\right)\Bigl)\Biggr]
    + \biggl(\frac{\alpha-\epsilon}{2S_2\epsilon}\biggl)\Biggl\{\Biggr[ 2 \gamma ^2 f_0^2 \epsilon ^2 t_c \bigl(m^2 \omega _0^2 t_c^2+\left(\gamma  t_c+m\right)^2 \bigl)\\
    &-2 f_0^2 m^2 \epsilon  t_c \omega _c\biggl( t_c^2 \left(\alpha ^2 \gamma +\gamma ^3 \omega _0^2+\epsilon \omega _c \left(\gamma ^2+m^2 \omega _0^2\right)-\gamma  m^2 \omega _0^2 \omega _c^2\right)-m^2 \left(\epsilon \omega _c-2 \gamma  m \epsilon  t_c \omega _c\right)\biggl)\Biggr]\\
    &+ \dfrac{\gamma   \epsilon\Delta_2}{m}    \left(\epsilon  \left(T_c+T_h\right) \left(\gamma ^2+m^2 \omega _c^2\right)+\alpha  \gamma ^2 \left(T_c-T_h\right)\right)\Biggl\}.
\end{split}   
\label{eq:Heat_full_SD}
\end{equation}
\end{widetext}

Based on the values of $P$ and $Q_h$, in Fig.~\ref{fig:Regimes_sd}, we have shown the parameter regimes in the parameter space $\omega_c-\epsilon$, for which the system functions as a heat engine or heat pump.
It is worth noting that the system is stable only when the correlation matrix $\boldsymbol{\Xi}'$ is positive definite. In our model, this stability condition is achieved only when $S_2 \le 0$, where $S_2$ is given in Eq.~\eqref{eq:S2_def}. For $S_2 > 0$, the system is unstable. This condition reduces to the known results in Ref.~\cite{park2016efficiency} in the limits $t_c \to 0$ and $\alpha \to 0$.
For small values of $\epsilon$, $Q_h$ remains positive even when $P$ is negative, indicating that the system cannot function as a heat pump or a heat engine (the green region in Fig.~\ref{fig:Regimes_sd}). Keeping the value of $\omega_c$ fixed, if we continue to load by increasing the value of $\epsilon$, there is a point ($ \epsilon = \epsilon_r$), at which $Q_h$ vanishes ($Q_h = 0$). This condition $\epsilon=\epsilon_r$ for all values $\omega_c$ defines the boundary line of the green regime along which $Q_h$ becomes zero. Beyond this point $\epsilon=\epsilon_r$, if $\epsilon$ is further increased, $Q_h$ becomes negative, reversing the direction of the heat flow and allowing the system to function as a heat pump (the blue region in Fig.~\ref{fig:Regimes_sd}).
This region in which the system works as a heat pump becomes narrower with an increase in the value of $\omega_c$. Additionally, as $t_c$ increases, this region initially expands, reaches a maximum area for an intermediate $t_c$ value [see Fig.~\ref{fig:Regimes_sd}(b)], and further decreases for larger values of $t_c$ [Fig.~\ref{fig:Regimes_sd}(c)].
With further increase in $\epsilon$ values, the system becomes unstable, since $S_2>0$, as discussed above. This region is denoted by grey color in Fig.~\ref{fig:Regimes_sd}. The system does not attain a steady state in this regime, making it unstable.

\begin{figure}
\centering
\includegraphics[width=\linewidth]{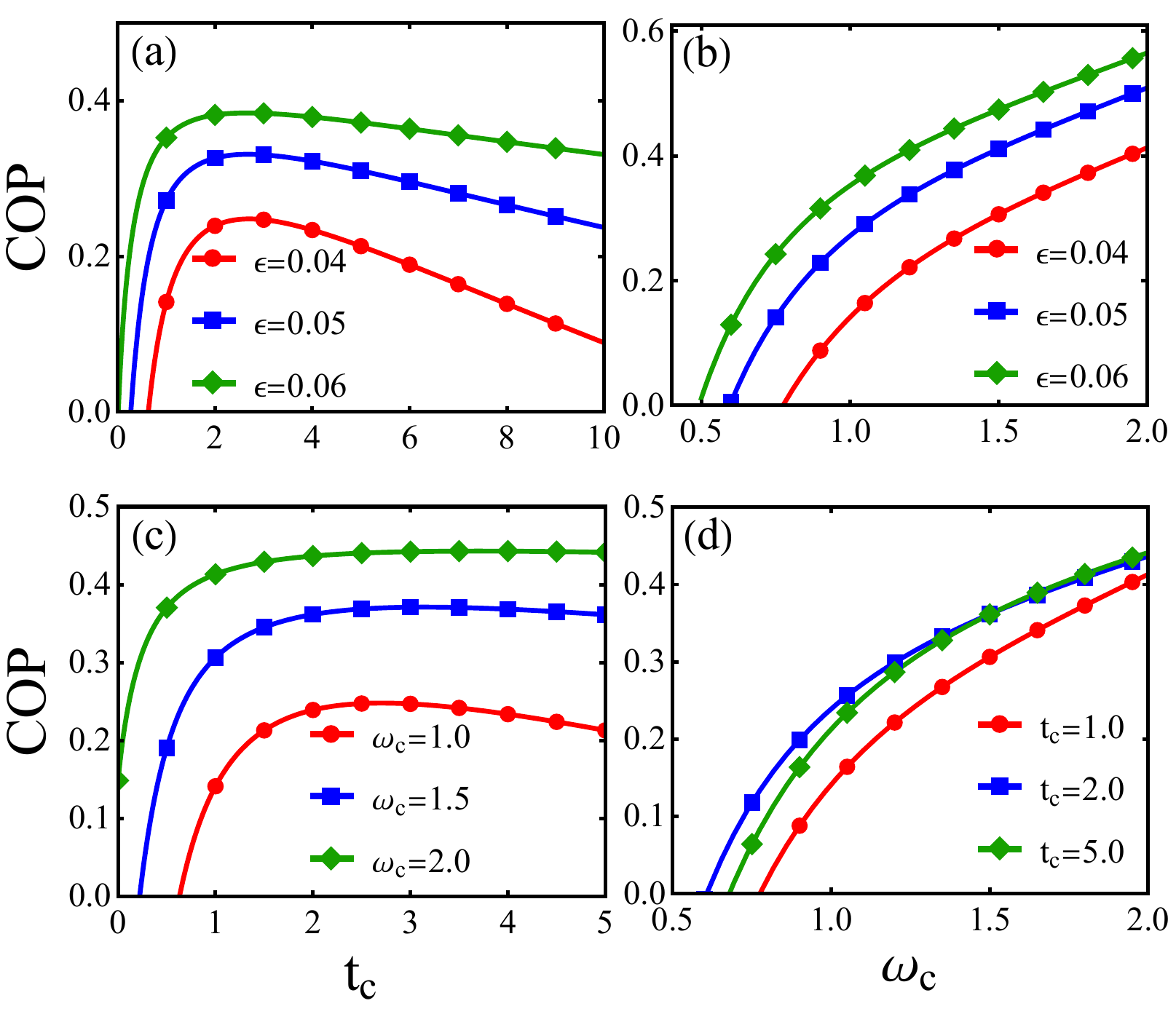}
\caption{The COP as a function of $t_c$ is shown in (a) for different values of $\epsilon$ and in (c) for different values of $\omega_c$. Similarly, the COP as a function of $\omega_c$ is shown in (b) for different values of $\epsilon$ and in (d) for different values of $t_c$. In (c) and (d), $\epsilon$=0.04is taken. The common parameters are  $m = f_0 = 1.0$, $\gamma=0.1$, $\alpha=0.75$, and $\omega_0=1.0$.}
\label{fig:COP_tc_wc_2nd}
\end{figure}

In order to quantify the performance of the heat pump in the blue regions of Fig.~\ref{fig:Regimes_sd}, we have exactly calculated the COP of the heat pump using Eq.~\eqref{eq:COP_main}, and it is given by
\begin{equation}
    \text{COP}=-\frac{\lambda_2}{\epsilon}-\frac{\alpha}{2\epsilon}+\frac{1}{2}.
    \label{eq:COP_2nd}
\end{equation}
Here, $\lambda_2$ is
\begin{widetext}
    \begin{equation}
        \lambda_2=\dfrac{\omega _c^2\left[2 \alpha  f_0^2 m \epsilon  t_c \left(m \omega _0^2 t_c^2+2 \gamma  t_c+m\right)-\gamma \Delta _2  \left[\left(T_c-T_h\right) \left(\epsilon ^2-\gamma ^2 \omega _0^2\right)-\alpha  \epsilon  \left(T_c+T_h\right)\right)+m \omega _0^2 t_c^2 \left(\epsilon +\gamma \omega _c\right)+2 \gamma  \epsilon  t_c+m \epsilon \right]}{\omega _c \left[\gamma  \Delta _2 m^2 \epsilon  \omega _c \left(T_c-T_h\right)-f_0^2 m^2 t_c \left(t_c \left(-2 \gamma  m \epsilon  \omega _c-t_c \left(\gamma  \left(\alpha ^2+\gamma  \epsilon  \omega _c\right)+\omega _0^2 \left(\gamma ^3+m^2 \omega _c \left(\gamma  \omega _c+\epsilon \right)\right)\right)\right)-m^2 \epsilon  \omega _c\right)\right]-S_4},
        \label{eq:lambda2}
    \end{equation}
with
\begin{equation}
    S_4=\gamma ^2 \left[-f_0^2 \epsilon  t_c \left(m^2 \omega _0^2 t_c^2+\left(\gamma  t_c+m\right)^2 \right)+\gamma  \Delta _2 (\alpha -\epsilon ) \left(T_c-T_h\right)\right].
\end{equation}
\end{widetext}
Next, we have plotted the COP as a function of $t_c$ for different values of $\epsilon$ and $\omega_c$ in Figs.~\ref{fig:COP_tc_wc_2nd} (a) and (c), and as a function of $\omega_c$ for different values of $\epsilon$ and $t_c$ in Figs.~\ref{fig:COP_tc_wc_2nd}(b) and (d), respectively. We observe that the COP increases with an increase in $\omega_c$ value, while it exhibits a non-monotonic dependence on $t_c$. It initially increases, attains a maximum value and then decreases with a further increase in the $t_c$ values. Furthermore, COP improves with increasing values of $\epsilon$.
Therefore, when the system is loaded in the same direction as the torque, both the activity and the magnetic field can enhance the performance of the heat pump. Since COP is an increasing function of $\omega_c$, an active magneto heat pump can be realized with high values of COP, provided that the system remains within the stable region.

For the vanishing activity limit, one can obtain $P$, $Q_h$, and COP by taking the limit $t_c \to 0$ in Eqs.~\eqref{eq:Power_FE}, \eqref{eq:Heat_full_SD}, and \eqref{eq:COP_2nd}, respectively.  The parameter regime where the system functions as a heat pump in this limit is significantly smaller compared to the active magneto-gyrator case. This indicates a limited tunability of the heat pump. Similarly, taking the $\omega_c \to 0$ limit of Eq.~\eqref{eq:COP_2nd} gives the COP of an active Brownian heat pump. However, in this limit, the regime in which the system functions as a heat pump remains very small, hence putting a limitation on the tunability of the parameters. Finally, from the analysis of our results, it is confirmed that in the limit $\omega_c \to 0$, both the efficiency and the coefficient of performance [Eqs.~\eqref{eq:Eff_1st}, ~\eqref{eq:COP_1st}, and ~\eqref{eq:COP_2nd}] do not have dependence on the duration of activity.


\section{CONCLUSIONS}\label{sec:summary}
In this work, we have analytically investigated the self-propulsion of a charged active particle confined in a two-dimensional asymmetric parabolic potential and simultaneously coupled to two heat baths kept at two different temperatures. Additionally, a magnetic field of constant amplitude is applied perpendicular to the plane of motion of the particle. Such a system is referred to as an active magneto-gyrator, and the particle exhibits a torque on the confining potential as long as there is a potential asymmetry and temperature gradient. We have exactly explored the thermodynamic performance characteristics of the system by applying a load force in a direction opposite as well as in the same direction of torque. It is observed that when the load is applied in a direction opposing the torque, the gyrator performs both as a heat engine and as a heat pump. While performing as an engine, the efficiency of the gyrator exhibits a non-monotonic dependence both on the duration of activity and the strength of the magnetic field. Unlike the case of a passive Brownian gyrator or Brownian magneto gyrator, interestingly, the efficiency of the active magneto gyrator is found to have no universal upper bound and can be brought to unity by tuning the system parameters. This is a clear indication of $100\%$ conversion of heat from the hot bath to mechanical work. However, this behavior is not feasible for a highly viscous medium, where the inertial effect is negligible. Further, when the gyrator is subjected to a viscoelastic bath characterized by a finite memory, for a short persistence of memory and fixed duration of activity, the gyrator is observed to achieve $100\%$ efficiency even for more than one value of memory time scale. 
Moreover, the duration of activity for which the efficiency becomes $100\%$ varies non-monotonically with short persistence of memory.
While performing as a heat pump, the coefficient of performance has a non-monotonic dependence on the duration of activity, whereas it decreases monotonically with an increase in the strength of the magnetic field. Hence, in this case, the magnetic field degrades the performance of the gyrator.

When the load is applied in the same direction as the torque, the system cannot function as a heat engine, as the extraction of work is not possible. However, it can operate as a heat pump in suitable parameter regimes depending on the direction of the heat flow.  
The coefficient of performance of the heat pump is an increasing function of the magnetic field strength. Hence, when the system is loaded in the same direction of torque, the magnetic field helps to enhance the performance. It is to be noted that we have analyzed the thermodynamic performance characteristics of active magneto- gyrator based on the calculations of steady-state averages of heat and extracted work, provided the system is in the stability regime. It would be further interesting to look at the fluctuating properties of the thermodynamic quantities in the stability regime. Finally, we believe that our analysis might be applicable to systems exhibiting circular motion in an anisotropic fluctuating field such as chiral colloidal microswimmers in parabolic potentials \cite{mancois2018two}, cicrular swimmers in a harmonic trap \cite{jahanshahi2017brownian}, active janus particles in complex plasma \cite{nosenko2020active}, etc. 

\section{Acknowledgement}
MS acknowledges the start-up grant from UGC, state plan fund from the University of Kerala, SERB-SURE grant (SUR/2022/000377), and CRG grant (CRG/2023/002026) from DST, Govt. of India for financial support.


\begin{thebibliography}{41}%
	\makeatletter
	\providecommand \@ifxundefined [1]{%
		\@ifx{#1\undefined}
	}%
	\providecommand \@ifnum [1]{%
		\ifnum #1\expandafter \@firstoftwo
		\else \expandafter \@secondoftwo
		\fi
	}%
	\providecommand \@ifx [1]{%
		\ifx #1\expandafter \@firstoftwo
		\else \expandafter \@secondoftwo
		\fi
	}%
	\providecommand \natexlab [1]{#1}%
	\providecommand \enquote  [1]{``#1''}%
	\providecommand \bibnamefont  [1]{#1}%
	\providecommand \bibfnamefont [1]{#1}%
	\providecommand \citenamefont [1]{#1}%
	\providecommand \href@noop [0]{\@secondoftwo}%
	\providecommand \href [0]{\begingroup \@sanitize@url \@href}%
	\providecommand \@href[1]{\@@startlink{#1}\@@href}%
	\providecommand \@@href[1]{\endgroup#1\@@endlink}%
	\providecommand \@sanitize@url [0]{\catcode `\\12\catcode `\$12\catcode `\&12\catcode `\#12\catcode `\^12\catcode `\_12\catcode `\%12\relax}%
	\providecommand \@@startlink[1]{}%
	\providecommand \@@endlink[0]{}%
	\providecommand \url  [0]{\begingroup\@sanitize@url \@url }%
	\providecommand \@url [1]{\endgroup\@href {#1}{\urlprefix }}%
	\providecommand \urlprefix  [0]{URL }%
	\providecommand \Eprint [0]{\href }%
	\providecommand \doibase [0]{https://doi.org/}%
	\providecommand \selectlanguage [0]{\@gobble}%
	\providecommand \bibinfo  [0]{\@secondoftwo}%
	\providecommand \bibfield  [0]{\@secondoftwo}%
	\providecommand \translation [1]{[#1]}%
	\providecommand \BibitemOpen [0]{}%
	\providecommand \bibitemStop [0]{}%
	\providecommand \bibitemNoStop [0]{.\EOS\space}%
	\providecommand \EOS [0]{\spacefactor3000\relax}%
	\providecommand \BibitemShut  [1]{\csname bibitem#1\endcsname}%
	\let\auto@bib@innerbib\@empty
	\bibitem [{\citenamefont {Filliger}\ and\ \citenamefont {Reimann}(2007)}]{filliger2007brownian}%
	\BibitemOpen
	\bibfield  {author} {\bibinfo {author} {\bibfnamefont {R.}~\bibnamefont {Filliger}}\ and\ \bibinfo {author} {\bibfnamefont {P.}~\bibnamefont {Reimann}},\ }\bibfield  {title} {\bibinfo {title} {Brownian gyrator: A minimal heat engine on the nanoscale},\ }\href {https://doi.org/10.1103/PhysRevLett.99.230602} {\bibfield  {journal} {\bibinfo  {journal} {Phys. Rev. Lett.}\ }\textbf {\bibinfo {volume} {99}},\ \bibinfo {pages} {230602} (\bibinfo {year} {2007})}\BibitemShut {NoStop}%
	\bibitem [{\citenamefont {Dotsenko}\ \emph {et~al.}(2013)\citenamefont {Dotsenko}, \citenamefont {Macio\l{}ek}, \citenamefont {Vasilyev},\ and\ \citenamefont {Oshanin}}]{dotsenko2013two}%
	\BibitemOpen
	\bibfield  {author} {\bibinfo {author} {\bibfnamefont {V.}~\bibnamefont {Dotsenko}}, \bibinfo {author} {\bibfnamefont {A.}~\bibnamefont {Macio\l{}ek}}, \bibinfo {author} {\bibfnamefont {O.}~\bibnamefont {Vasilyev}},\ and\ \bibinfo {author} {\bibfnamefont {G.}~\bibnamefont {Oshanin}},\ }\bibfield  {title} {\bibinfo {title} {Two-temperature langevin dynamics in a parabolic potential},\ }\href {https://doi.org/10.1103/PhysRevE.87.062130} {\bibfield  {journal} {\bibinfo  {journal} {Phys. Rev. E}\ }\textbf {\bibinfo {volume} {87}},\ \bibinfo {pages} {062130} (\bibinfo {year} {2013})}\BibitemShut {NoStop}%
	\bibitem [{\citenamefont {Mancois}\ \emph {et~al.}(2018)\citenamefont {Mancois}, \citenamefont {Marcos}, \citenamefont {Viot},\ and\ \citenamefont {Wilkowski}}]{mancois2018two}%
	\BibitemOpen
	\bibfield  {author} {\bibinfo {author} {\bibfnamefont {V.}~\bibnamefont {Mancois}}, \bibinfo {author} {\bibfnamefont {B.}~\bibnamefont {Marcos}}, \bibinfo {author} {\bibfnamefont {P.}~\bibnamefont {Viot}},\ and\ \bibinfo {author} {\bibfnamefont {D.}~\bibnamefont {Wilkowski}},\ }\bibfield  {title} {\bibinfo {title} {Two-temperature brownian dynamics of a particle in a confining potential},\ }\href {https://doi.org/10.1103/PhysRevE.97.052121} {\bibfield  {journal} {\bibinfo  {journal} {Phys. Rev. E}\ }\textbf {\bibinfo {volume} {97}},\ \bibinfo {pages} {052121} (\bibinfo {year} {2018})}\BibitemShut {NoStop}%
	\bibitem [{\citenamefont {Chiang}\ \emph {et~al.}(2017)\citenamefont {Chiang}, \citenamefont {Lee}, \citenamefont {Lai},\ and\ \citenamefont {Chen}}]{Chiang2017electrical}%
	\BibitemOpen
	\bibfield  {author} {\bibinfo {author} {\bibfnamefont {K.-H.}\ \bibnamefont {Chiang}}, \bibinfo {author} {\bibfnamefont {C.-L.}\ \bibnamefont {Lee}}, \bibinfo {author} {\bibfnamefont {P.-Y.}\ \bibnamefont {Lai}},\ and\ \bibinfo {author} {\bibfnamefont {Y.-F.}\ \bibnamefont {Chen}},\ }\bibfield  {title} {\bibinfo {title} {Electrical autonomous brownian gyrator},\ }\href {https://doi.org/10.1103/PhysRevE.96.032123} {\bibfield  {journal} {\bibinfo  {journal} {Phys. Rev. E}\ }\textbf {\bibinfo {volume} {96}},\ \bibinfo {pages} {032123} (\bibinfo {year} {2017})}\BibitemShut {NoStop}%
	\bibitem [{\citenamefont {Bae}\ \emph {et~al.}(2021)\citenamefont {Bae}, \citenamefont {Lee}, \citenamefont {Kim},\ and\ \citenamefont {Jeong}}]{bae2021inertial}%
	\BibitemOpen
	\bibfield  {author} {\bibinfo {author} {\bibfnamefont {Y.}~\bibnamefont {Bae}}, \bibinfo {author} {\bibfnamefont {S.}~\bibnamefont {Lee}}, \bibinfo {author} {\bibfnamefont {J.}~\bibnamefont {Kim}},\ and\ \bibinfo {author} {\bibfnamefont {H.}~\bibnamefont {Jeong}},\ }\bibfield  {title} {\bibinfo {title} {Inertial effects on the brownian gyrator},\ }\href {https://doi.org/10.1103/PhysRevE.103.032148} {\bibfield  {journal} {\bibinfo  {journal} {Phys. Rev. E}\ }\textbf {\bibinfo {volume} {103}},\ \bibinfo {pages} {032148} (\bibinfo {year} {2021})}\BibitemShut {NoStop}%
	\bibitem [{\citenamefont {Argun}\ \emph {et~al.}(2017)\citenamefont {Argun}, \citenamefont {Soni}, \citenamefont {Dabelow}, \citenamefont {Bo}, \citenamefont {Pesce}, \citenamefont {Eichhorn},\ and\ \citenamefont {Volpe}}]{Argun2017Experimental}%
	\BibitemOpen
	\bibfield  {author} {\bibinfo {author} {\bibfnamefont {A.}~\bibnamefont {Argun}}, \bibinfo {author} {\bibfnamefont {J.}~\bibnamefont {Soni}}, \bibinfo {author} {\bibfnamefont {L.}~\bibnamefont {Dabelow}}, \bibinfo {author} {\bibfnamefont {S.}~\bibnamefont {Bo}}, \bibinfo {author} {\bibfnamefont {G.}~\bibnamefont {Pesce}}, \bibinfo {author} {\bibfnamefont {R.}~\bibnamefont {Eichhorn}},\ and\ \bibinfo {author} {\bibfnamefont {G.}~\bibnamefont {Volpe}},\ }\bibfield  {title} {\bibinfo {title} {Experimental realization of a minimal microscopic heat engine},\ }\href {https://doi.org/10.1103/PhysRevE.96.052106} {\bibfield  {journal} {\bibinfo  {journal} {Phys. Rev. E}\ }\textbf {\bibinfo {volume} {96}},\ \bibinfo {pages} {052106} (\bibinfo {year} {2017})}\BibitemShut {NoStop}%
	\bibitem [{\citenamefont {Chang}\ \emph {et~al.}(2021)\citenamefont {Chang}, \citenamefont {Lee}, \citenamefont {Lai},\ and\ \citenamefont {Chen}}]{Chang2021autonomous}%
	\BibitemOpen
	\bibfield  {author} {\bibinfo {author} {\bibfnamefont {H.}~\bibnamefont {Chang}}, \bibinfo {author} {\bibfnamefont {C.-L.}\ \bibnamefont {Lee}}, \bibinfo {author} {\bibfnamefont {P.-Y.}\ \bibnamefont {Lai}},\ and\ \bibinfo {author} {\bibfnamefont {Y.-F.}\ \bibnamefont {Chen}},\ }\bibfield  {title} {\bibinfo {title} {Autonomous brownian gyrators: A study on gyrating characteristics},\ }\href {https://doi.org/10.1103/PhysRevE.103.022128} {\bibfield  {journal} {\bibinfo  {journal} {Phys. Rev. E}\ }\textbf {\bibinfo {volume} {103}},\ \bibinfo {pages} {022128} (\bibinfo {year} {2021})}\BibitemShut {NoStop}%
	\bibitem [{\citenamefont {Baldassarri}\ \emph {et~al.}(2020)\citenamefont {Baldassarri}, \citenamefont {Puglisi},\ and\ \citenamefont {Sesta}}]{baldassarri2020engineered}%
	\BibitemOpen
	\bibfield  {author} {\bibinfo {author} {\bibfnamefont {A.}~\bibnamefont {Baldassarri}}, \bibinfo {author} {\bibfnamefont {A.}~\bibnamefont {Puglisi}},\ and\ \bibinfo {author} {\bibfnamefont {L.}~\bibnamefont {Sesta}},\ }\bibfield  {title} {\bibinfo {title} {Engineered swift equilibration of a brownian gyrator},\ }\href {https://doi.org/10.1103/PhysRevE.102.030105} {\bibfield  {journal} {\bibinfo  {journal} {Phys. Rev. E}\ }\textbf {\bibinfo {volume} {102}},\ \bibinfo {pages} {030105} (\bibinfo {year} {2020})}\BibitemShut {NoStop}%
	\bibitem [{\citenamefont {Cerasoli}\ \emph {et~al.}(2021)\citenamefont {Cerasoli}, \citenamefont {Dotsenko}, \citenamefont {Oshanin},\ and\ \citenamefont {Rondoni}}]{cerasoli2021time}%
	\BibitemOpen
	\bibfield  {author} {\bibinfo {author} {\bibfnamefont {S.}~\bibnamefont {Cerasoli}}, \bibinfo {author} {\bibfnamefont {V.}~\bibnamefont {Dotsenko}}, \bibinfo {author} {\bibfnamefont {G.}~\bibnamefont {Oshanin}},\ and\ \bibinfo {author} {\bibfnamefont {L.}~\bibnamefont {Rondoni}},\ }\bibfield  {title} {\bibinfo {title} {Time-dependence of the effective temperatures of a two-dimensional brownian gyrator with cold and hot components},\ }\href {https://doi.org/10.1088/1751-8121/abe0d6} {\bibfield  {journal} {\bibinfo  {journal} {Journal of Physics A: Mathematical and Theoretical}\ }\textbf {\bibinfo {volume} {54}},\ \bibinfo {pages} {105002} (\bibinfo {year} {2021})}\BibitemShut {NoStop}%
	\bibitem [{\citenamefont {Fogedby}\ and\ \citenamefont {Imparato}(2017)}]{fogedby2017minimal}%
	\BibitemOpen
	\bibfield  {author} {\bibinfo {author} {\bibfnamefont {H.~C.}\ \bibnamefont {Fogedby}}\ and\ \bibinfo {author} {\bibfnamefont {A.}~\bibnamefont {Imparato}},\ }\bibfield  {title} {\bibinfo {title} {A minimal model of an autonomous thermal motor},\ }\href {https://doi.org/10.1209/0295-5075/119/50007} {\bibfield  {journal} {\bibinfo  {journal} {Europhysics Letters}\ }\textbf {\bibinfo {volume} {119}},\ \bibinfo {pages} {50007} (\bibinfo {year} {2017})}\BibitemShut {NoStop}%
	\bibitem [{\citenamefont {Cerasoli}\ \emph {et~al.}(2022)\citenamefont {Cerasoli}, \citenamefont {Ciliberto}, \citenamefont {Marinari}, \citenamefont {Oshanin}, \citenamefont {Peliti},\ and\ \citenamefont {Rondoni}}]{cerasoli2022spectral}%
	\BibitemOpen
	\bibfield  {author} {\bibinfo {author} {\bibfnamefont {S.}~\bibnamefont {Cerasoli}}, \bibinfo {author} {\bibfnamefont {S.}~\bibnamefont {Ciliberto}}, \bibinfo {author} {\bibfnamefont {E.}~\bibnamefont {Marinari}}, \bibinfo {author} {\bibfnamefont {G.}~\bibnamefont {Oshanin}}, \bibinfo {author} {\bibfnamefont {L.}~\bibnamefont {Peliti}},\ and\ \bibinfo {author} {\bibfnamefont {L.}~\bibnamefont {Rondoni}},\ }\bibfield  {title} {\bibinfo {title} {Spectral fingerprints of nonequilibrium dynamics: The case of a brownian gyrator},\ }\href {https://doi.org/10.1103/PhysRevE.106.014137} {\bibfield  {journal} {\bibinfo  {journal} {Phys. Rev. E}\ }\textbf {\bibinfo {volume} {106}},\ \bibinfo {pages} {014137} (\bibinfo {year} {2022})}\BibitemShut {NoStop}%
	\bibitem [{\citenamefont {Murashita}\ and\ \citenamefont {Esposito}(2016)}]{murashita2016overdamped}%
	\BibitemOpen
	\bibfield  {author} {\bibinfo {author} {\bibfnamefont {Y.}~\bibnamefont {Murashita}}\ and\ \bibinfo {author} {\bibfnamefont {M.}~\bibnamefont {Esposito}},\ }\bibfield  {title} {\bibinfo {title} {Overdamped stochastic thermodynamics with multiple reservoirs},\ }\href {https://doi.org/10.1103/PhysRevE.94.062148} {\bibfield  {journal} {\bibinfo  {journal} {Phys. Rev. E}\ }\textbf {\bibinfo {volume} {94}},\ \bibinfo {pages} {062148} (\bibinfo {year} {2016})}\BibitemShut {NoStop}%
	\bibitem [{\citenamefont {Nascimento}\ and\ \citenamefont {Morgado}(2020)}]{nascimento2020memory}%
	\BibitemOpen
	\bibfield  {author} {\bibinfo {author} {\bibfnamefont {E.~S.}\ \bibnamefont {Nascimento}}\ and\ \bibinfo {author} {\bibfnamefont {W.~A.~M.}\ \bibnamefont {Morgado}},\ }\bibfield  {title} {\bibinfo {title} {Memory effects on two-dimensional overdamped brownian dynamics},\ }\href {https://doi.org/10.1088/1751-8121/ab5e2b} {\bibfield  {journal} {\bibinfo  {journal} {J. Phys. A: Math. Theor.}\ }\textbf {\bibinfo {volume} {53}},\ \bibinfo {pages} {065001} (\bibinfo {year} {2020})}\BibitemShut {NoStop}%
	\bibitem [{\citenamefont {Van~den Broeck}\ \emph {et~al.}(2004)\citenamefont {Van~den Broeck}, \citenamefont {Kawai},\ and\ \citenamefont {Meurs}}]{broeck2004micro}%
	\BibitemOpen
	\bibfield  {author} {\bibinfo {author} {\bibfnamefont {C.}~\bibnamefont {Van~den Broeck}}, \bibinfo {author} {\bibfnamefont {R.}~\bibnamefont {Kawai}},\ and\ \bibinfo {author} {\bibfnamefont {P.}~\bibnamefont {Meurs}},\ }\bibfield  {title} {\bibinfo {title} {Microscopic analysis of a thermal brownian motor},\ }\href {https://doi.org/10.1103/PhysRevLett.93.090601} {\bibfield  {journal} {\bibinfo  {journal} {Phys. Rev. Lett.}\ }\textbf {\bibinfo {volume} {93}},\ \bibinfo {pages} {090601} (\bibinfo {year} {2004})}\BibitemShut {NoStop}%
	\bibitem [{\citenamefont {Bérut}\ \emph {et~al.}(2016)\citenamefont {Bérut}, \citenamefont {Imparato}, \citenamefont {Petrosyan},\ and\ \citenamefont {Ciliberto}}]{berut2016role}%
	\BibitemOpen
	\bibfield  {author} {\bibinfo {author} {\bibfnamefont {A.}~\bibnamefont {Bérut}}, \bibinfo {author} {\bibfnamefont {A.}~\bibnamefont {Imparato}}, \bibinfo {author} {\bibfnamefont {A.}~\bibnamefont {Petrosyan}},\ and\ \bibinfo {author} {\bibfnamefont {S.}~\bibnamefont {Ciliberto}},\ }\bibfield  {title} {\bibinfo {title} {The role of coupling on the statistical properties of the energy fluxes between stochastic systems at different temperatures},\ }\href {https://doi.org/10.1088/1742-5468/2016/05/054002} {\bibfield  {journal} {\bibinfo  {journal} {Journal of Statistical Mechanics: Theory and Experiment}\ }\textbf {\bibinfo {volume} {2016}},\ \bibinfo {pages} {054002} (\bibinfo {year} {2016})}\BibitemShut {NoStop}%
	\bibitem [{\citenamefont {Crisanti}\ \emph {et~al.}(2012)\citenamefont {Crisanti}, \citenamefont {Puglisi},\ and\ \citenamefont {Villamaina}}]{crisanti2012nonequilibrium}%
	\BibitemOpen
	\bibfield  {author} {\bibinfo {author} {\bibfnamefont {A.}~\bibnamefont {Crisanti}}, \bibinfo {author} {\bibfnamefont {A.}~\bibnamefont {Puglisi}},\ and\ \bibinfo {author} {\bibfnamefont {D.}~\bibnamefont {Villamaina}},\ }\bibfield  {title} {\bibinfo {title} {Nonequilibrium and information: The role of cross correlations},\ }\href {https://doi.org/10.1103/PhysRevE.85.061127} {\bibfield  {journal} {\bibinfo  {journal} {Phys. Rev. E}\ }\textbf {\bibinfo {volume} {85}},\ \bibinfo {pages} {061127} (\bibinfo {year} {2012})}\BibitemShut {NoStop}%
	\bibitem [{\citenamefont {Park}\ \emph {et~al.}(2016)\citenamefont {Park}, \citenamefont {Chun},\ and\ \citenamefont {Noh}}]{park2016efficiency}%
	\BibitemOpen
	\bibfield  {author} {\bibinfo {author} {\bibfnamefont {J.-M.}\ \bibnamefont {Park}}, \bibinfo {author} {\bibfnamefont {H.-M.}\ \bibnamefont {Chun}},\ and\ \bibinfo {author} {\bibfnamefont {J.~D.}\ \bibnamefont {Noh}},\ }\bibfield  {title} {\bibinfo {title} {Efficiency at maximum power and efficiency fluctuations in a linear brownian heat-engine model},\ }\href {https://doi.org/10.1103/PhysRevE.94.012127} {\bibfield  {journal} {\bibinfo  {journal} {Phys. Rev. E}\ }\textbf {\bibinfo {volume} {94}},\ \bibinfo {pages} {012127} (\bibinfo {year} {2016})}\BibitemShut {NoStop}%
	\bibitem [{\citenamefont {Abdoli}\ \emph {et~al.}(2022)\citenamefont {Abdoli}, \citenamefont {Wittmann}, \citenamefont {Brader}, \citenamefont {Sommer}, \citenamefont {L{\"o}wen},\ and\ \citenamefont {Sharma}}]{abdoli2022tunable}%
	\BibitemOpen
	\bibfield  {author} {\bibinfo {author} {\bibfnamefont {I.}~\bibnamefont {Abdoli}}, \bibinfo {author} {\bibfnamefont {R.}~\bibnamefont {Wittmann}}, \bibinfo {author} {\bibfnamefont {J.~M.}\ \bibnamefont {Brader}}, \bibinfo {author} {\bibfnamefont {J.-U.}\ \bibnamefont {Sommer}}, \bibinfo {author} {\bibfnamefont {H.}~\bibnamefont {L{\"o}wen}},\ and\ \bibinfo {author} {\bibfnamefont {A.}~\bibnamefont {Sharma}},\ }\bibfield  {title} {\bibinfo {title} {Tunable brownian magneto heat pump},\ }\href {https://doi.org/10.1038/s41598-022-17584-3} {\bibfield  {journal} {\bibinfo  {journal} {Sci. Rep.}\ }\textbf {\bibinfo {volume} {12}},\ \bibinfo {pages} {13405} (\bibinfo {year} {2022})}\BibitemShut {NoStop}%
	\bibitem [{\citenamefont {Mart{\'i}nez}\ \emph {et~al.}(2016)\citenamefont {Mart{\'i}nez}, \citenamefont {Rold{\'a}n}, \citenamefont {Dinis}, \citenamefont {Petrov}, \citenamefont {Parrondo},\ and\ \citenamefont {Rica}}]{martinez2016carnot}%
	\BibitemOpen
	\bibfield  {author} {\bibinfo {author} {\bibfnamefont {I.~A.}\ \bibnamefont {Mart{\'i}nez}}, \bibinfo {author} {\bibfnamefont {{\'E}.}~\bibnamefont {Rold{\'a}n}}, \bibinfo {author} {\bibfnamefont {L.}~\bibnamefont {Dinis}}, \bibinfo {author} {\bibfnamefont {D.}~\bibnamefont {Petrov}}, \bibinfo {author} {\bibfnamefont {J.~M.~R.}\ \bibnamefont {Parrondo}},\ and\ \bibinfo {author} {\bibfnamefont {R.~A.}\ \bibnamefont {Rica}},\ }\bibfield  {title} {\bibinfo {title} {Brownian carnot engine},\ }\href {https://doi.org/10.1038/nphys3518} {\bibfield  {journal} {\bibinfo  {journal} {Nature Physics}\ }\textbf {\bibinfo {volume} {12}},\ \bibinfo {pages} {67} (\bibinfo {year} {2016})}\BibitemShut {NoStop}%
	\bibitem [{\citenamefont {Frim}\ and\ \citenamefont {DeWeese}(2022)}]{adam2022browniancarnot}%
	\BibitemOpen
	\bibfield  {author} {\bibinfo {author} {\bibfnamefont {A.~G.}\ \bibnamefont {Frim}}\ and\ \bibinfo {author} {\bibfnamefont {M.~R.}\ \bibnamefont {DeWeese}},\ }\bibfield  {title} {\bibinfo {title} {Optimal finite-time brownian carnot engine},\ }\href {https://doi.org/10.1103/PhysRevE.105.L052103} {\bibfield  {journal} {\bibinfo  {journal} {Phys. Rev. E}\ }\textbf {\bibinfo {volume} {105}},\ \bibinfo {pages} {L052103} (\bibinfo {year} {2022})}\BibitemShut {NoStop}%
	\bibitem [{\citenamefont {Abdoli}\ \emph {et~al.}(2020)\citenamefont {Abdoli}, \citenamefont {Kalz}, \citenamefont {Vuijk}, \citenamefont {Wittmann}, \citenamefont {Sommer}, \citenamefont {Brader},\ and\ \citenamefont {Sharma}}]{abdoli2020correlations}%
	\BibitemOpen
	\bibfield  {author} {\bibinfo {author} {\bibfnamefont {I.}~\bibnamefont {Abdoli}}, \bibinfo {author} {\bibfnamefont {E.}~\bibnamefont {Kalz}}, \bibinfo {author} {\bibfnamefont {H.~D.}\ \bibnamefont {Vuijk}}, \bibinfo {author} {\bibfnamefont {R.}~\bibnamefont {Wittmann}}, \bibinfo {author} {\bibfnamefont {J.-U.}\ \bibnamefont {Sommer}}, \bibinfo {author} {\bibfnamefont {J.~M.}\ \bibnamefont {Brader}},\ and\ \bibinfo {author} {\bibfnamefont {A.}~\bibnamefont {Sharma}},\ }\bibfield  {title} {\bibinfo {title} {Correlations in multithermostat brownian systems with lorentz force},\ }\href {https://doi.org/10.1088/1367-2630/abb43d} {\bibfield  {journal} {\bibinfo  {journal} {New J. Phys.}\ }\textbf {\bibinfo {volume} {22}},\ \bibinfo {pages} {093057} (\bibinfo {year} {2020})}\BibitemShut {NoStop}%
	\bibitem [{\citenamefont {Kumari}\ \emph {et~al.}(2024)\citenamefont {Kumari}, \citenamefont {Samsuzzaman}, \citenamefont {Saha},\ and\ \citenamefont {Lahiri}}]{kumari2024Stochastic}%
	\BibitemOpen
	\bibfield  {author} {\bibinfo {author} {\bibfnamefont {A.}~\bibnamefont {Kumari}}, \bibinfo {author} {\bibfnamefont {M.}~\bibnamefont {Samsuzzaman}}, \bibinfo {author} {\bibfnamefont {A.}~\bibnamefont {Saha}},\ and\ \bibinfo {author} {\bibfnamefont {S.}~\bibnamefont {Lahiri}},\ }\bibfield  {title} {\bibinfo {title} {Stochastic heat engine using multiple interacting active particles},\ }\href {https://doi.org/https://doi.org/10.1016/j.physa.2024.129575} {\bibfield  {journal} {\bibinfo  {journal} {Physica A: Statistical Mechanics and its Applications}\ }\textbf {\bibinfo {volume} {636}},\ \bibinfo {pages} {129575} (\bibinfo {year} {2024})}\BibitemShut {NoStop}%
	\bibitem [{\citenamefont {Saha}\ and\ \citenamefont {Marathe}(2019)}]{saha2019stochastic}%
	\BibitemOpen
	\bibfield  {author} {\bibinfo {author} {\bibfnamefont {A.}~\bibnamefont {Saha}}\ and\ \bibinfo {author} {\bibfnamefont {R.}~\bibnamefont {Marathe}},\ }\bibfield  {title} {\bibinfo {title} {Stochastic work extraction in a colloidal heat engine in the presence of colored noise},\ }\href {https://doi.org/10.1088/1742-5468/ab39d4} {\bibfield  {journal} {\bibinfo  {journal} {Journal of Statistical Mechanics: Theory and Experiment}\ }\textbf {\bibinfo {volume} {2019}},\ \bibinfo {pages} {094012} (\bibinfo {year} {2019})}\BibitemShut {NoStop}%
	\bibitem [{\citenamefont {Pietzonka}\ and\ \citenamefont {Seifert}(2017)}]{Pietzonka2018Entropy}%
	\BibitemOpen
	\bibfield  {author} {\bibinfo {author} {\bibfnamefont {P.}~\bibnamefont {Pietzonka}}\ and\ \bibinfo {author} {\bibfnamefont {U.}~\bibnamefont {Seifert}},\ }\bibfield  {title} {\bibinfo {title} {Entropy production of active particles and for particles in active baths},\ }\href {https://doi.org/10.1088/1751-8121/aa91b9} {\bibfield  {journal} {\bibinfo  {journal} {Journal of Physics A: Mathematical and Theoretical}\ }\textbf {\bibinfo {volume} {51}},\ \bibinfo {pages} {01LT01} (\bibinfo {year} {2017})}\BibitemShut {NoStop}%
	\bibitem [{\citenamefont {Mandal}\ \emph {et~al.}(2017)\citenamefont {Mandal}, \citenamefont {Klymko},\ and\ \citenamefont {DeWeese}}]{mandal2017entropy}%
	\BibitemOpen
	\bibfield  {author} {\bibinfo {author} {\bibfnamefont {D.}~\bibnamefont {Mandal}}, \bibinfo {author} {\bibfnamefont {K.}~\bibnamefont {Klymko}},\ and\ \bibinfo {author} {\bibfnamefont {M.~R.}\ \bibnamefont {DeWeese}},\ }\bibfield  {title} {\bibinfo {title} {Entropy production and fluctuation theorems for active matter},\ }\href {https://doi.org/10.1103/PhysRevLett.119.258001} {\bibfield  {journal} {\bibinfo  {journal} {Phys. Rev. Lett.}\ }\textbf {\bibinfo {volume} {119}},\ \bibinfo {pages} {258001} (\bibinfo {year} {2017})}\BibitemShut {NoStop}%
	\bibitem [{\citenamefont {Datta}\ \emph {et~al.}(2022)\citenamefont {Datta}, \citenamefont {Pietzonka},\ and\ \citenamefont {Barato}}]{datta2022second}%
	\BibitemOpen
	\bibfield  {author} {\bibinfo {author} {\bibfnamefont {A.}~\bibnamefont {Datta}}, \bibinfo {author} {\bibfnamefont {P.}~\bibnamefont {Pietzonka}},\ and\ \bibinfo {author} {\bibfnamefont {A.~C.}\ \bibnamefont {Barato}},\ }\bibfield  {title} {\bibinfo {title} {Second law for active heat engines},\ }\href {https://doi.org/10.1103/PhysRevX.12.031034} {\bibfield  {journal} {\bibinfo  {journal} {Phys. Rev. X}\ }\textbf {\bibinfo {volume} {12}},\ \bibinfo {pages} {031034} (\bibinfo {year} {2022})}\BibitemShut {NoStop}%
	\bibitem [{\citenamefont {Saha}\ \emph {et~al.}(2018)\citenamefont {Saha}, \citenamefont {Marathe}, \citenamefont {Pal},\ and\ \citenamefont {Jayannavar}}]{saha2018stochastic}%
	\BibitemOpen
	\bibfield  {author} {\bibinfo {author} {\bibfnamefont {A.}~\bibnamefont {Saha}}, \bibinfo {author} {\bibfnamefont {R.}~\bibnamefont {Marathe}}, \bibinfo {author} {\bibfnamefont {P.~S.}\ \bibnamefont {Pal}},\ and\ \bibinfo {author} {\bibfnamefont {A.~M.}\ \bibnamefont {Jayannavar}},\ }\bibfield  {title} {\bibinfo {title} {Stochastic heat engine powered by active dissipation},\ }\href {https://doi.org/10.1088/1742-5468/aae84a} {\bibfield  {journal} {\bibinfo  {journal} {Journal of Statistical Mechanics: Theory and Experiment}\ }\textbf {\bibinfo {volume} {2018}},\ \bibinfo {pages} {113203} (\bibinfo {year} {2018})}\BibitemShut {NoStop}%
	\bibitem [{\citenamefont {Kumari}\ \emph {et~al.}(2020)\citenamefont {Kumari}, \citenamefont {Pal}, \citenamefont {Saha},\ and\ \citenamefont {Lahiri}}]{kumari2020stochastic}%
	\BibitemOpen
	\bibfield  {author} {\bibinfo {author} {\bibfnamefont {A.}~\bibnamefont {Kumari}}, \bibinfo {author} {\bibfnamefont {P.}~\bibnamefont {Pal}}, \bibinfo {author} {\bibfnamefont {A.}~\bibnamefont {Saha}},\ and\ \bibinfo {author} {\bibfnamefont {S.}~\bibnamefont {Lahiri}},\ }\bibfield  {title} {\bibinfo {title} {Stochastic heat engine using an active particle},\ }\bibfield  {journal} {\bibinfo  {journal} {Physical Review E}\ }\textbf {\bibinfo {volume} {101}},\ \href {https://doi.org/10.1103/PhysRevE.101.032109} {10.1103/PhysRevE.101.032109} (\bibinfo {year} {2020}),\ \bibinfo {note} {cited by: 40; All Open Access, Green Open Access}\BibitemShut {NoStop}%
	\bibitem [{\citenamefont {Pietzonka}\ \emph {et~al.}(2019)\citenamefont {Pietzonka}, \citenamefont {Fodor}, \citenamefont {Lohrmann}, \citenamefont {Cates},\ and\ \citenamefont {Seifert}}]{pietzonka2019Autonomous}%
	\BibitemOpen
	\bibfield  {author} {\bibinfo {author} {\bibfnamefont {P.}~\bibnamefont {Pietzonka}}, \bibinfo {author} {\bibfnamefont {E.}~\bibnamefont {Fodor}}, \bibinfo {author} {\bibfnamefont {C.}~\bibnamefont {Lohrmann}}, \bibinfo {author} {\bibfnamefont {M.~E.}\ \bibnamefont {Cates}},\ and\ \bibinfo {author} {\bibfnamefont {U.}~\bibnamefont {Seifert}},\ }\bibfield  {title} {\bibinfo {title} {Autonomous engines driven by active matter: Energetics and design principles},\ }\href {https://doi.org/10.1103/PhysRevX.9.041032} {\bibfield  {journal} {\bibinfo  {journal} {Phys. Rev. X}\ }\textbf {\bibinfo {volume} {9}},\ \bibinfo {pages} {041032} (\bibinfo {year} {2019})}\BibitemShut {NoStop}%
	\bibitem [{\citenamefont {Hecht}\ \emph {et~al.}(2022)\citenamefont {Hecht}, \citenamefont {Mandal}, \citenamefont {L\"owen},\ and\ \citenamefont {Liebchen}}]{HechtActive2022}%
	\BibitemOpen
	\bibfield  {author} {\bibinfo {author} {\bibfnamefont {L.}~\bibnamefont {Hecht}}, \bibinfo {author} {\bibfnamefont {S.}~\bibnamefont {Mandal}}, \bibinfo {author} {\bibfnamefont {H.}~\bibnamefont {L\"owen}},\ and\ \bibinfo {author} {\bibfnamefont {B.}~\bibnamefont {Liebchen}},\ }\bibfield  {title} {\bibinfo {title} {Active refrigerators powered by inertia},\ }\href {https://doi.org/10.1103/PhysRevLett.129.178001} {\bibfield  {journal} {\bibinfo  {journal} {Phys. Rev. Lett.}\ }\textbf {\bibinfo {volume} {129}},\ \bibinfo {pages} {178001} (\bibinfo {year} {2022})}\BibitemShut {NoStop}%
	\bibitem [{\citenamefont {Muhsin}\ \emph {et~al.}(2025)\citenamefont {Muhsin}, \citenamefont {Adersh},\ and\ \citenamefont {Sahoo}}]{Muhsin2025Active}%
	\BibitemOpen
	\bibfield  {author} {\bibinfo {author} {\bibfnamefont {M.}~\bibnamefont {Muhsin}}, \bibinfo {author} {\bibfnamefont {F.}~\bibnamefont {Adersh}},\ and\ \bibinfo {author} {\bibfnamefont {M.}~\bibnamefont {Sahoo}},\ }\bibfield  {title} {\bibinfo {title} {Active magneto gyrator: Memory-induced trapped diamagnetism},\ }\href {https://doi.org/10.1103/PhysRevE.111.015411} {\bibfield  {journal} {\bibinfo  {journal} {Phys. Rev. E}\ }\textbf {\bibinfo {volume} {111}},\ \bibinfo {pages} {015411} (\bibinfo {year} {2025})}\BibitemShut {NoStop}%
	\bibitem [{\citenamefont {Muhsin}\ and\ \citenamefont {Sahoo}(2022)}]{muhsin2022inertial}%
	\BibitemOpen
	\bibfield  {author} {\bibinfo {author} {\bibfnamefont {M.}~\bibnamefont {Muhsin}}\ and\ \bibinfo {author} {\bibfnamefont {M.}~\bibnamefont {Sahoo}},\ }\bibfield  {title} {\bibinfo {title} {Inertial active ornstein-uhlenbeck particle in the presence of a magnetic field},\ }\href {https://doi.org/10.1103/PhysRevE.106.014605} {\bibfield  {journal} {\bibinfo  {journal} {Phys. Rev. E}\ }\textbf {\bibinfo {volume} {106}},\ \bibinfo {pages} {014605} (\bibinfo {year} {2022})}\BibitemShut {NoStop}%
	\bibitem [{\citenamefont {Noushad}\ \emph {et~al.}(2021)\citenamefont {Noushad}, \citenamefont {Shajahan},\ and\ \citenamefont {Sahoo}}]{arsha2021velocity}%
	\BibitemOpen
	\bibfield  {author} {\bibinfo {author} {\bibfnamefont {A.}~\bibnamefont {Noushad}}, \bibinfo {author} {\bibfnamefont {S.}~\bibnamefont {Shajahan}},\ and\ \bibinfo {author} {\bibfnamefont {M.}~\bibnamefont {Sahoo}},\ }\bibfield  {title} {\bibinfo {title} {{V}elocity auto correlation function of a confined {B}rownian particle},\ }\href {https://doi.org/10.1140/epjb/s10051-021-00217-5} {\bibfield  {journal} {\bibinfo  {journal} {Eur. Phys. J. B}\ }\textbf {\bibinfo {volume} {94}},\ \bibinfo {pages} {202} (\bibinfo {year} {2021})}\BibitemShut {NoStop}%
	\bibitem [{\citenamefont {Caprini}\ and\ \citenamefont {Marini Bettolo~Marconi}(2021)}]{caprini2021inertial}%
	\BibitemOpen
	\bibfield  {author} {\bibinfo {author} {\bibfnamefont {L.}~\bibnamefont {Caprini}}\ and\ \bibinfo {author} {\bibfnamefont {U.}~\bibnamefont {Marini Bettolo~Marconi}},\ }\bibfield  {title} {\bibinfo {title} {Inertial self-propelled particles},\ }\href {https://doi.org/10.1063/5.0030940} {\bibfield  {journal} {\bibinfo  {journal} {J. Chem. Phys.}\ }\textbf {\bibinfo {volume} {154}},\ \bibinfo {pages} {024902} (\bibinfo {year} {2021})}\BibitemShut {NoStop}%
	\bibitem [{\citenamefont {Sekimoto}(1998)}]{sekimoto1997langevin}%
	\BibitemOpen
	\bibfield  {author} {\bibinfo {author} {\bibfnamefont {K.}~\bibnamefont {Sekimoto}},\ }\bibfield  {title} {\bibinfo {title} {Langevin equation and thermodynamics},\ }\href {https://doi.org/10.1143/PTPS.130.17} {\bibfield  {journal} {\bibinfo  {journal} {Progress of Theoretical Physics Supplement}\ }\textbf {\bibinfo {volume} {130}},\ \bibinfo {pages} {17} (\bibinfo {year} {1998})}\BibitemShut {NoStop}%
	\bibitem [{\citenamefont {Sekimoto}()}]{sekimoto2010stochastic}%
	\BibitemOpen
	\bibfield  {author} {\bibinfo {author} {\bibfnamefont {K.}~\bibnamefont {Sekimoto}},\ }\href@noop {} {\emph {\bibinfo {title} {Stochastic energetics}}},\ \bibinfo {series} {Lecture notes in physics}\ No.\ \bibinfo {number} {799}\ (\bibinfo  {publisher} {Springer})\BibitemShut {NoStop}%
	\bibitem [{\citenamefont {Qi}\ \emph {et~al.}(2022)\citenamefont {Qi}, \citenamefont {Chen}, \citenamefont {Ge}, \citenamefont {Yang},\ and\ \citenamefont {Feng}}]{qi2022thermal}%
	\BibitemOpen
	\bibfield  {author} {\bibinfo {author} {\bibfnamefont {C.}~\bibnamefont {Qi}}, \bibinfo {author} {\bibfnamefont {L.}~\bibnamefont {Chen}}, \bibinfo {author} {\bibfnamefont {Y.}~\bibnamefont {Ge}}, \bibinfo {author} {\bibfnamefont {W.}~\bibnamefont {Yang}},\ and\ \bibinfo {author} {\bibfnamefont {H.}~\bibnamefont {Feng}},\ }\bibfield  {title} {\bibinfo {title} {Thermal brownian heat pump with external and internal irreversibilities},\ }\href {https://doi.org/10.1140/epjp/s13360-022-03287-1} {\bibfield  {journal} {\bibinfo  {journal} {The European Physical Journal Plus}\ }\textbf {\bibinfo {volume} {137}},\ \bibinfo {pages} {1079} (\bibinfo {year} {2022})}\BibitemShut {NoStop}%
	\bibitem [{\citenamefont {Chen}\ \emph {et~al.}(2024)\citenamefont {Chen}, \citenamefont {Qi}, \citenamefont {Ge},\ and\ \citenamefont {Feng}}]{chen2024three}%
	\BibitemOpen
	\bibfield  {author} {\bibinfo {author} {\bibfnamefont {L.}~\bibnamefont {Chen}}, \bibinfo {author} {\bibfnamefont {C.}~\bibnamefont {Qi}}, \bibinfo {author} {\bibfnamefont {Y.}~\bibnamefont {Ge}},\ and\ \bibinfo {author} {\bibfnamefont {H.}~\bibnamefont {Feng}},\ }\bibfield  {title} {\bibinfo {title} {Three-heat-reservoir thermal brownian heat pump and its performance limits},\ }\href {https://doi.org/10.1016/j.csite.2024.104224} {\bibfield  {journal} {\bibinfo  {journal} {Case Studies in Thermal Engineering}\ }\textbf {\bibinfo {volume} {56}},\ \bibinfo {pages} {104224} (\bibinfo {year} {2024})}\BibitemShut {NoStop}%
	\bibitem [{\citenamefont {{VAN KAMPEN}}(2007)}]{van2007stochastic}%
	\BibitemOpen
	\bibfield  {author} {\bibinfo {author} {\bibfnamefont {N.}~\bibnamefont {{VAN KAMPEN}}},\ }\bibfield  {title} {\bibinfo {title} {Chapter viii - the fokker–planck equation},\ }in\ \href {https://doi.org/https://doi.org/10.1016/B978-044452965-7/50011-8} {\emph {\bibinfo {booktitle} {Stochastic Processes in Physics and Chemistry (Third Edition)}}},\ \bibinfo {series and number} {North-Holland Personal Library},\ \bibinfo {editor} {edited by\ \bibinfo {editor} {\bibfnamefont {N.}~\bibnamefont {{VAN KAMPEN}}}}\ (\bibinfo  {publisher} {Elsevier},\ \bibinfo {address} {Amsterdam},\ \bibinfo {year} {2007})\ \bibinfo {edition} {third edition}\ ed.,\ pp.\ \bibinfo {pages} {193--218}\BibitemShut {NoStop}%
	\bibitem [{\citenamefont {Jahanshahi}\ \emph {et~al.}(2017)\citenamefont {Jahanshahi}, \citenamefont {L\"owen},\ and\ \citenamefont {ten Hagen}}]{jahanshahi2017brownian}%
	\BibitemOpen
	\bibfield  {author} {\bibinfo {author} {\bibfnamefont {S.}~\bibnamefont {Jahanshahi}}, \bibinfo {author} {\bibfnamefont {H.}~\bibnamefont {L\"owen}},\ and\ \bibinfo {author} {\bibfnamefont {B.}~\bibnamefont {ten Hagen}},\ }\bibfield  {title} {\bibinfo {title} {Brownian motion of a circle swimmer in a harmonic trap},\ }\href {https://doi.org/10.1103/PhysRevE.95.022606} {\bibfield  {journal} {\bibinfo  {journal} {Phys. Rev. E}\ }\textbf {\bibinfo {volume} {95}},\ \bibinfo {pages} {022606} (\bibinfo {year} {2017})}\BibitemShut {NoStop}%
	\bibitem [{\citenamefont {Nosenko}\ \emph {et~al.}(2020)\citenamefont {Nosenko}, \citenamefont {Luoni}, \citenamefont {Kaouk}, \citenamefont {Rubin-Zuzic},\ and\ \citenamefont {Thomas}}]{nosenko2020active}%
	\BibitemOpen
	\bibfield  {author} {\bibinfo {author} {\bibfnamefont {V.}~\bibnamefont {Nosenko}}, \bibinfo {author} {\bibfnamefont {F.}~\bibnamefont {Luoni}}, \bibinfo {author} {\bibfnamefont {A.}~\bibnamefont {Kaouk}}, \bibinfo {author} {\bibfnamefont {M.}~\bibnamefont {Rubin-Zuzic}},\ and\ \bibinfo {author} {\bibfnamefont {H.}~\bibnamefont {Thomas}},\ }\bibfield  {title} {\bibinfo {title} {Active janus particles in a complex plasma},\ }\href {https://doi.org/10.1103/PhysRevResearch.2.033226} {\bibfield  {journal} {\bibinfo  {journal} {Phys. Rev. Res.}\ }\textbf {\bibinfo {volume} {2}},\ \bibinfo {pages} {033226} (\bibinfo {year} {2020})}\BibitemShut {NoStop}%
\end{thebibliography}
\end{document}